\documentclass[prd,11pt,secnumarabic,eqsecnum,nofootinbib,raggedbottom]{revtex4-1}

\usepackage{amsmath}    
\usepackage{amssymb}
\usepackage{mathtools}
\usepackage{graphicx}   
\usepackage{verbatim}   
\usepackage{color}      
\usepackage{subfigure}  
\usepackage{hyperref}   
\usepackage{mathrsfs}
\usepackage{slashed}
\usepackage{verbatim}
\usepackage{latexsym}
\usepackage{euscript}
\usepackage{amsfonts}
\usepackage{verbatim}
\usepackage[]{array}
\usepackage[]{mathrsfs}
\usepackage{amsxtra}
\usepackage{braket}
\usepackage{cancel}
\usepackage{units}
\usepackage{ulem} 
\usepackage{hyperref}
\usepackage{mdframed}
\usepackage{changepage}

\newcommand\numberthis{\addtocounter{equation}{1}\tag{\theequation}}

\begin{document}

\title{On the particle content of MHS theory}\thanks{Preprint number: ZTF-EP-24-08}

\author{Maro Cvitan}
\affiliation{Department of Physics, Faculty of Science, University of Zagreb, Zagreb, Croatia.}
\email{mcvitan@phy.hr}
\author{Predrag Dominis Prester}
\affiliation{University of Rijeka, Faculty of Mathematics, Rijeka, Croatia}
\email{pprester@math.uniri.hr}
\author{Stefano Giaccari}
\affiliation{Istituto Nazionale di Ricerca Metrologica, Torino, Italy
}
\email{s.giaccari@inrim.it}
\author{Mateo Paulišić}
\affiliation{University of Rijeka, Faculty of Physics, Rijeka, Croatia}
\email{mateo.paulisic@uniri.hr}
\author{Ivan Vuković}
\affiliation{Vienna, Austria}
\email{vukovic.ac@gmail.com}

\begin{abstract}
The Moyal-Higher-Spin (MHS) formalism, involving fields dependent on spacetime and auxiliary coordinates, is an approach to studying higher spin (HS)-like models. To determine the particle content of the MHS model of the Yang-Mills type, we calculate the quartic Casimir operator for on-shell MHS fields, finding it generally non-vanishing, indicative of infinite/continuous-spin degrees of freedom. We propose an on-shell basis for these infinite/continuous-spin states. Additionally, we analyse the content of a massive MHS model.
\end{abstract}

\maketitle

\section{Introduction}
\label{linear_and_content}

There are both purely theoretical and phenomenological reasons for the construction of consistent higher spin (spin $s>2$) quantum field theories (QFTs) in Minkowski background, which remains a still elusive goal. Such theories are expected to contain an infinite tower of higher spin fields/particles, a property which could ensure a better (softer) UV behaviour and, in this way, avoid some of the problems present in standard QFTs (e.g., Landau pole). If the HS fields are described by totally symmetric Lorentz tensor spacetime fields $\phi_{\mu_1\cdots\mu_s}(x)$, the infinite tower can be packed into a field on a $2d$-dimensional domain by using an auxiliary space with coordinates $u = {u_\mu,\mu=0,\ldots,d-1}$ transforming as a Lorentz vector, in the following way
\begin{equation} \label{gfhsf}
\phi(x,u) = \sum_{s=0}^\infty \phi^{\mu_1\cdots\mu_s}(x)\, u_{\mu_1} \cdots u_{\mu_s}
\end{equation}
We refer to fields defined in the $2d$-dimensional space spanned by $x$ and $u$ as master fields. Usually, master fields are used just as a formal construct to write equations for free HS spacetime fields in a compact way, in particular without any requirements of convergence of the infinite series in (\ref{gfhsf}). We observe here that requiring convergence in (\ref{gfhsf}) automatically implies constraints on spacetime fields which means that they cannot be independent. However, recently in the literature, there appeared constructions in which master fields {\it are} used as fundamental objects in the off-shell descriptions \cite{Cvitan:2021yvf,Cvitan:2021qvm,Cvitan:2022wzf,Schuster:2013pta}. Here we shall be interested in the approaches in which master fields are required to be square-integrable in the auxiliary space
\begin{equation}
\int d^d u\, \phi(x,u)^\dagger\, \phi(x,u) < \infty\,,
\end{equation} an example being given by the Moyal-Higher-Spin (MHS) gauge theories \cite{Bonora:2018uwx,Bonora:2018eot, Bonora:2018ggh, Bonora:2018mqg, Bonora:2020aqp, Cvitan:2021qvm, Cvitan:2022wzf}. In this case, a Taylor expansion (\ref{gfhsf}) is not much of use as the spacetime fields defined in this way cannot be treated as independent degrees of freedom. Consequently, one cannot read off the particle content of the theory by plugging (\ref{gfhsf}) into the linearised equations of motion, so one has to use different methods to understand how Lorentz and Poincaré groups are represented. 
These representations can be studied in general for any dependence on $u$, as well as in a particular basis in the auxiliary space $u$.
The Poincaré transformations, in general, act on master fields in the following way
\begin{equation} \label{mfPt}
\phi'_r(x,u) = \sum_s \mathscr{D}(\Lambda)_r{}^s\, \phi_s(\Lambda^{-1}x - \zeta\,, \Lambda^{-1}u)\,,
\end{equation}
where $r$ and $s$ represent a finite number of Lorentz indices and $\mathscr{D}(\Lambda)$ is some standard finite-dimensional IRREP of the Lorentz group (which for a scalar master field is trivial). 
It was noted in \cite{Schuster:2013pxj} that one can use an infinite-dimensional representation of the Lorentz group induced by the transformation of the auxiliary coordinate $u$ in (\ref{mfPt}). It was further noted in \cite{Cvitan:2021qvm} that the restriction of master field dependence on $u$ to the linear vector space $L_2(\mathbb{R}^d)$ (square-integrable functions over the auxiliary space) is both \textit{infinite-dimensional} and \textit{unitary}.
The explicit expression of the matrix representation, written in the basis of $d$-dimensional Hermite functions, was presented in \cite{Cvitan:2024xmn}.
What remained to be analysed is the perturbative spectrum (i.e.\ particle content of a free theory) of a master field that is square integrable in the auxiliary space, in terms of IRREP of the Poincaré group by the Wigner construction, and this is what we do in the present paper. To avoid unnecessary complications, we shall demonstrate the construction on the example of the master field gauge potential in the YM theory, both for the massless and massive cases.

We begin by reviewing the MHS formalism and then display several approaches to analysing the theory's spacetime content. The first one, a Taylor expansion in the auxiliary space, is conventional and enables a direct comparison to the standard higher spin models. However, as mentioned, it is not consistent with the requirement of the square integrability of the master fields. 
The second approach is an expansion in terms of the orthonormal basis of functions (for discrete bases, orthonormality is defined using Kroneckers and for continuous bases using Dirac delta functions) in the auxiliary space \cite{Cvitan:2021yvf}. 
We employ modes given by products of Hermite functions (momentum independent). Such modes furnish an infinite-dimensional representation of the Lorentz group (\cite{Cvitan:2021qvm}, \cite{Cvitan:2024xmn}). 
We also employ modes that are momentum-dependent and are solutions to the differential equations posed by the little group generators.

By analysing the polarisation structure of the solutions of the linear equations of motion of the MHSYM model (linearised around Minkowski vacuum) expanded in terms of Hermite functions, we learn about supported helicities and obtain an indication that, in general, we can have a non-vanishing value of the quartic Casimir. In addition, by analysing the polarisation structure, now in terms of functions which are solutions to the differential equations posed by the little group generators, we find a complete on-shell (momentum-dependent) basis which enables us to read off the particle content of the solutions. In the appendix~\ref{AppMassive}, we analyse the particle content of a massive MHS master field, working directly on the space of one-particle states.
\section{Master field Yang-Mills theories}

As a working example, let us use the generalisation of Yang-Mills theories defined on the flat Minkowski background and write here just the basic aspects and equations important for the context of the present paper. An explicit example of such a class of theories is the MHS gauge theory developed in \cite{Cvitan:2021yvf}. While the theory can be defined for an arbitrary number of spacetime dimensions $d$, we shall focus mostly on the case $d=4$. The basic object is the gauge potential master field $h_a(x,u)$, where $x = \{x^\mu, \mu=0,\ldots d-1\}$ are spacetime coordinates, $u = \{u_\mu, \mu=0,\ldots d-1\}$ are (dimensionless) coordinates in the auxiliary space transforming as a Lorentz (covariant) vector, and $a = \{0,\ldots,d-1\}$ is a frame index. Infinitesimal gauge transformations are given by
\begin{equation}\label{hepsilon}
\delta h_a(x,u) = \partial_a \varepsilon(x,u) + O(h)\,,
\end{equation}
where the gauge parameter $\varepsilon(x,u)$ is a generic (infinitesimal) function satisfying some appropriate boundary conditions. Note that we assume here the simplest case of the Maxwell-like theories based on the $U(1)$ internal gauge symmetry. The formalism can be naturally extended to non-commutative internal gauge groups. 
However, the structure of the terms linear in the gauge potential depends on the theory in question and is not important for this paper. In the example of the MHS theory, YM algebraic structure is defined by the Moyal product, which introduces a non-commutative structure between the spacetime and the auxiliary space. This structure guarantees not only good classical behaviour but also enables one to formally use standard quantisation methods. 
Generalising the Yang-Mills procedure, one defines the master field strength
\begin{equation}
F_{ab}(x,u) = \partial^x_a h_b(x,u) - \partial^x_b h_a(x,u) + O(h^2)\,,
\end{equation}
and the YM action
\begin{equation}\label{YMaction}
S[h,\psi] = S_{\mathrm{ym}}[h] + S_{\mathrm{matt}}[h,\psi] \,,
\end{equation}
where $\psi$ denotes (minimally coupled) matter in the form of master fields and/or ordinary spacetime fields, and
\begin{equation}
\label{ymact}
S_{\mathrm{ym}} = - \frac{1}{4} \int d^dx\, d^du\, F^{ab}(x,u) \, F_{ab}(x,u)\,.
\end{equation}
The equations of motion for the gauge master field are
\begin{align}
\label{ymeom}
\Box_x h_a - \partial^x_a \partial^x_b h^b + m_h^2 h_a + O(h^2) 
= \frac{\delta S'_{\mathrm{matt}}}{\delta h^a} \,,
\end{align}
where $S'_\mathrm{matt}$ is part of $S_{\mathrm{matt}}$ that does not include the mass term, possibly generated by the Higgs mechanism. The energy in the linear approximation is given by
\begin{equation}
\label{enlapp}
E \approx \frac{1}{2} \int d\mathbf{x}\int d^du
\left( \sum_j F_{0j}(x,u)^2 + \sum_{j<k} F_{jk}(x,u)^2 + m_h^2 \left( h_0(x,u)^2+ \sum_i h_i(x,u)^2\right) \right) + U_{\mathrm{matt}}\,,
\end{equation}
where $j,k \in {1,\ldots,d-1}$. From (\ref{ymeom}) and (\ref{enlapp}), we can conclude that the theory is classically consistent, at least in the perturbative domain: the vacuum $h_a = 0$ is stable, and the energy is positive-definite and finite.

Furthermore, there are no problems with ghosts in perturbative expansions. To understand this, 
let us first rewrite the MHS gauge field transformation under Poincaré transformations in the following way
\begin{equation}
h'_a(x,u) = \Lambda_a{}^b\, \left( D(\Lambda)\, h_b \right)(\Lambda^{-1}x - \zeta, u)\,,
\end{equation}
\begin{equation}\label{auxgaug}
\left( D(\Lambda)\, h_b \right)(x, u) = h_b(x, \Lambda^{-1}u)\,.
\end{equation}
$D(\Lambda)$ defines an infinite dimensional linear representation of the Lorentz group on $L_2(\mathbb{R}^d)$, the vector space of square-integrable functions over the auxiliary space. Moreover, this representation is {\it unitary} with respect to the standard inner product on $L_2(\mathbb{R}^d)$
\begin{equation}
\langle \Psi | \Phi \rangle = \int d^d u\, \Psi(u)^\dagger\, \Phi(u)\,.
\end{equation}
The proof is
\begin{equation} \label{unproof}
\int d^du\, \left(D(\Lambda)\Psi(u)\right)^\dagger D(\Lambda)\Phi(u) = \int d^d u\, \Psi^\dagger(u\Lambda)\Phi(u\Lambda)  = \int d^du\, \Psi^\dagger(u)\Phi(u) \,,
\end{equation}
where we used $d^d(u\Lambda) = d^d u$.
As $\Phi$ and $\Psi$ are arbitrary elements of $L_2(\mathbb{R}^d)$, it follows from (\ref{unproof}) that
\begin{equation}
D(\Lambda)^\dagger = D(\Lambda)^{-1}\,.
\end{equation}

The unitarity of $D(\Lambda)$ guarantees that the master gauge symmetry is large enough to deal with all physically unacceptable modes --- i.e.\ (\ref{hepsilon}) is sufficient to remove them. In fact, it was shown \cite{Bonora:2018ggh} that the theory can be formally quantised by using the standard methods of YM theory.

The next task is to investigate the spectrum (particle content) of such theories. In what follows, we shall assume that the master fields, as functions of auxiliary coordinates $u$, are essentially restricted only by the condition of square integrability.

\section{Spacetime fields}
\subsection{Taylor expansion}
\label{taylor_content}
In general, when dealing with higher spin fields, it is useful to pack a complete tower of higher spin fields into a single structure by using an auxiliary Lorentz vector as a bookkeeping device (see e.g.\ \cite{Bengtsson:2008mw}). 

By reversing this logic, it may appear that the master potential should be understood, from the purely spacetime viewpoint, as an infinite collection of HS fields obtained from  
\begin{equation} \label{hscte}
h_a(x,u) = \sum_{n=0}^\infty h_a^{(n)\mu_1\cdots\mu_n}(x)\, u_{\mu_1} \ldots u_{\mu_n} \,,
\end{equation}
where we use a Latin index for the master field and Greek indices for variables of expansion.
The coefficients in the expansion are spacetime fields that are Lorentz tensors of rank $n+1$, symmetric in their $n$ (Greek) indices, and which by (\ref{ymeom}) in the massless case satisfy Maxwell-type equations of motion 
\begin{equation} \label{ymsteom}
\Box h_a^{(n)\mu_1\cdots\mu_n} - \partial_a \partial^b h_b^{(n)\mu_1\cdots\mu_n} + O(h^2) = 0\,,
\end{equation}
while from (\ref{hepsilon}), we can deduce that the gauge transformations, obtained from expanding the gauge parameter as in (\ref{hscte}), are of the form
\begin{equation}\label{linearised_gauge_transf}
\delta_\varepsilon h_a^{(n)\mu_1\cdots\mu_n}(x) = \partial_a \varepsilon^{\mu_1\cdots\mu_n}(x) + O(h)\,.
\end{equation}

While there is a priori nothing wrong with the manifestly Lorentz covariant power expansion (\ref{hscte}), it cannot by itself be used for the purpose of uncovering the spectrum of physical excitations (particle spectrum) of the theory. This is because when substituting (\ref{hscte}) into the action (\ref{ymact}) or the energy (\ref{enlapp}) and organising terms by the order of $u_\mu$, we encounter divergent integrations over the auxiliary space $u$ of the form $\int d^du \, u_{\mu_1} \ldots u_{\mu_n}$ at each order $n$. Therefore, the requirement of square integrability in the auxiliary space forces us to abandon (\ref{hscte}) as a useful means.

\subsection{Orthogonal functions expansion}
An alternative to the Taylor expansion above comes from relaxing the notion of how Lorentz covariance is to be achieved and giving priority to the fact that integrals over the auxiliary space should be finite. We can then use a complete orthonormal set of functions in the auxiliary space $\{f_r(u)\}$, indexed by some formal parameter $r$ to expand the master potential as
\begin{equation} \label{mpone}
h_a(x,u) = \sum_r h_a^{(r)}(x)\, f_r(u)\,,
\end{equation}
where 
\begin{equation} \label{fuonb}
\int d^d u\, f_r(u)\, f_s(u) = \delta_{rs} \;.
\end{equation}
Using such an expansion, one arrives at the purely spacetime off-shell description with the free (quadratic) part of the Lagrangian  given by
\begin{equation} \label{Msa0}
S_0[h] = - \frac{1}{4 g_{\mathrm{ym}}^2} \sum_{r,s} \int d^dx\, \big(\partial_a h_b^{(r)} - \partial_b h_a^{(r)}\big)
\eta^{ac} \eta^{bd} \delta_{rs} \big(\partial_c h_d^{(s)} - \partial_d h_c^{(s)}\big) \;.
\end{equation}
On the linear level, the gauge symmetry acts on spacetime fields 
${h}_a^{(r)}(x)$ as
\begin{equation}
\delta_\varepsilon {h}_a^{(r)}(x) = \partial_a \varepsilon^{(r)}(x) + O(h)\,,
\end{equation}
where $\varepsilon^{(r)}(x)$ are obtained from the master gauge parameter $\varepsilon(x,u)$ by expanding as in (\ref{mpone}). 
One comes to the description in terms of an (infinite) set of Maxwell-like fields. This form shows explicitly the absence of physically non-acceptable modes (ghosts and/or negative energy modes) in the spectrum of the free theory.

The appearance of the Kronecker in (\ref{fuonb}), however, leads to another effect. The positive definite product of two basis functions in (\ref{fuonb}) might seem in disagreement with the condition of Lorentz covariance since intuition usually leads us to expect the Minkowski metric on the right-hand side of equations such as (\ref{fuonb}) if Lorentz covariance is to be achieved. However, as we have shown in the previous section, the Lorentz covariance is still present in the form of a unitary infinite-dimensional representation of the Lorentz group acting on the index $r$.
One particularly convenient choice for the orthonormal basis of functions are multi-dimensional Hermite functions that we have used in \cite{Cvitan:2024xmn}. We repeat the definition here with modifications due to the fact that the auxiliary space variables are $u_a$, not $u^a$.
$H_{n}(u)$ are the Hermite polynomials
\begin{equation}
H_{n}(u)=(-1)^{n} e^{u^{2}} \frac{d^{n}}{d u^{n}} e^{-u^{2}}
\,,
\end{equation}
where the index $n$ can attain arbitrary non-negative integer values. Hermite functions are defined as
\begin{equation}
f_{n}(u) = \frac{1}{\sqrt{2^n n! \sqrt{\pi}}}e^{-\frac{u^2}{2}} H_{n}(u)\,.
\end{equation}
The multi-dimensional Hermite function that we will use for the expansion in the auxiliary space is defined as
\begin{equation}
f_{n_0\cdots n_{d-1}}(u) = f_{n_0}(u_0)\cdots f_{n_{d-1}}(u_{d-1})\,.
\end{equation}
They satisfy the orthonormality condition
\begin{equation}\label{orthonormality_lower}
\int du_0\cdots du_{d-1}\,f_{n_0\cdots n_{d-1}}(u)f_{m_0\cdots m_{d-1}}(u) = \delta_{n_0}^{m_0}\cdots\delta_{n_{d-1}}^{m_{d-1}}\,.
\end{equation}
The MHS potential $h_a(x,u) \equiv h_a(x^b,u_c)$ is now expanded as
\begin{equation}\label{HermiteExpansionPotential}
h_a(x,u)=\sum_{\{n\}=0}^\infty h_a^{n_0\cdots n_{d-1}}(x)f_{n_0\cdots n_{d-1}}(u)\,.
\end{equation}
Following the transformation property 
\begin{equation}
\label{MHSunderLorentzPassive}
{h}^{\prime}_a(x^\prime,{u}^\prime) = \Lambda_a{}^{b}h_b(x,u)\,,
\end{equation} we can deduce the rules for Lorentz transformations of the component spacetime fields $ h_a^{n_0\cdots n_{d-1}}(x)$
\begin{equation}\label{LorentzActiveTransformation}
{h}_a^{\prime}(x,{u}) = \Lambda_a{}^{b}h_b(\Lambda^{-1}x,u\Lambda)
\end{equation}
and we can expand both sides of the equation in the Hermite basis
\begin{equation}
\sum_{\{n\}=0}^\infty h_a^{\prime n_0\cdots n_{d-1}}(x)f_{n_0\cdots n_{d-1}}(u) = \Lambda_a{}^{b}\sum_{\{m\}=0}^\infty h_b^{m_0\cdots m_{d-1}}(\Lambda^{-1}x)f_{m_0\cdots m_{d-1}}(u \Lambda)\,.
\end{equation}
Due to (\ref{orthonormality_lower}) we can multiply both sides with $f_{r_0\cdots r_d}(u)$, integrate over the auxiliary space, and conclude
\begin{align}
h_a^{\prime r_0\cdots r_{d-1}}(x)= \Lambda_a{}^{b}\sum_{\{m\}=0}^\infty L^{r_0\cdots r_{d-1}}_{m_0\cdots m_{d-1}}(\Lambda) h_b^{m_0 \cdots m_{d-1}}(\Lambda^{-1}x)\label{ha_under_Loretnz}\,,
\end{align}
where 
\begin{equation}\label{representation_for_casimir}
L^{r_0\cdots r_{d-1}}_{m_0\cdots m_{d-1}}(\Lambda)  = \int du_0\cdots du_{d-1}\,f_{r_0\cdots r_{d-1}}(u)f_{m_0\cdots m_{d-1}}(u\Lambda)
\end{equation}
is a representation matrix of the Lorentz group in the space of Hermite functions.

We have explicitly constructed the representation matrices in \cite{Cvitan:2024xmn}. 
Since here, the variables of integration are auxiliary space coordinates with lower Lorentz indices, while in \cite{Cvitan:2024xmn} we worked with variables with upper Lorentz indices,  
(\ref{representation_for_casimir}) differs slightly from the formula (3.11) of  \cite{Cvitan:2024xmn}. We now adapt the representation matrices to the current situation. For simplicity, we restrict to two dimensions. From  \cite{Cvitan:2024xmn} we have
\begin{equation}
D^{m_0m_1}_{n_0n_1}(\Lambda)=\int du^0du^1\, f_{m_0}(u^0)f_{m_1}(u^1)f_{n_0}((\Lambda^{-1}u)^0)f_{n_1}((\Lambda^{-1}u)^1)\,,
\end{equation}
while explicitly in (\ref{representation_for_casimir}) we have
\begin{equation}
L^{m_0m_1}_{n_0n_1}(\Lambda)=\int du_0du_1\, f_{m_0}(u_0)f_{m_1}(u_1)f_{n_0}((u\Lambda)_0)f_{n_1}((u\Lambda)_1)\,.
\end{equation}
In the mostly plus signature ($\eta^{00}=-1$, $\eta^{ii}=1$) that we are using, we can re-express
\begin{equation}
u_0 = - u^0,\quad u_1 = u^1,\quad (u\Lambda)_0 = - (u\Lambda)^0, \quad (u\Lambda)_1 = (u\Lambda)^1
\end{equation}
and further we realise
\begin{equation}
(u\Lambda)_\mu = u_\nu\Lambda^\nu{}_\mu,\quad (u\Lambda)^\mu=u_\nu\Lambda^\nu{}^\mu =u^\nu\Lambda_\nu{}^\mu =  (\Lambda^{-1})^\mu{}_\nu u^\nu = (\Lambda^{-1}u)^\mu\,.
\end{equation}
With the property of Hermite functions $f_n(-u) = (-1)^nf_n(u)$, we can finally relate
\begin{align}\label{representation_matrices_sign_change}
L^{m_0m_1}_{n_0n_1}(\Lambda)=(-1)^{n_0+m_0}D^{m_0m_1}_{n_0n_1}(\Lambda)\,.
\end{align}
With these conventions, the prefactor $(-1)^{n_0+m_0}$ does not depend on the number of spacetime dimensions. The representation matrices $D^{m_0m_1}_{n_0n_1}(\Lambda)$ can be found in \cite{Cvitan:2024xmn}.

For further convenience, we explicitly write down the generators of the Lorentz group in $d=4$ in the infinite-dimensional representation over Hermite functions, adapted to the purposes of this paper (here we also make the operators Hermitean so the rotation generators $J_i$ differ from those in \cite{Cvitan:2024xmn} by a global factor of $-i$, while the boost generators $K_i$ differ by a global factor of $i$).
\begin{align}\label{generators_for_Hermite_MHS1}
\hspace{-5mm}K_1{}_{n_0n_1n_2n_3}^{m_0m_1m_2m_3} = i\delta_{-n_0 + n_1+n_2+n_3}^{-m_0+m_1+m_2+m_3}\delta_{n_2}^{m_2}\delta_{n_3}^{m_3}\left(\delta^{m_1}_{n_1+1}\sqrt{(n_0+1)(n_1+1)} - \delta^{m_1}_{n_1-1}\sqrt{n_1n_0}\right)\\
\hspace{-5mm}K_2{}_{n_0n_1n_2n_3}^{m_0m_1m_2m_3} = i\delta_{-n_0 + n_1+n_2+n_3}^{-m_0+m_1+m_2+m_3}\delta_{n_1}^{m_1}\delta_{n_3}^{m_3}\left(\delta^{m_2}_{n_2+1}\sqrt{(n_0+1)(n_2+1)} - \delta^{m_2}_{n_2-1}\sqrt{n_2n_0}\right)\\
\hspace{-5mm}K_3{}_{n_0n_1n_2n_3}^{m_0m_1m_2m_3} = i\delta_{-n_0 + n_1+n_2+n_3}^{-m_0+m_1+m_2+m_3}\delta_{n_1}^{m_1}\delta_{n_2}^{m_2}\left(\delta^{m_3}_{n_3+1}\sqrt{(n_0+1)(n_3+1)} - \delta^{m_3}_{n_3-1}\sqrt{n_3n_0}\right)
\end{align}
\begin{align}
\hspace{-5mm}J_1{}^{m_0m_1m_2m_3}_{n_0n_1n_2n_3} = i\delta^{-m_0+m_1+m_2+m_3}_{-n_0+n_1+n_2+n_3}\delta^{m_1}_{n_1}\delta^{m_0}_{n_0}\left(\delta^{m_2}_{n_2+1}\sqrt{(n_2+1)n_3}-\delta^{m_2}_{n_2-1}\sqrt{n_2(n_3+1)}\right)\\
\hspace{-5mm}J_2{}^{m_0m_1m_2m_3}_{n_0n_1n_2n_3} = i\delta^{-m_0+m_1+m_2+m_3}_{-n_0+n_1+n_2+n_3}\delta^{m_2}_{n_2}\delta^{m_0}_{n_0}\left(\delta^{m_3}_{n_3+1}\sqrt{(n_3+1)n_1}-\delta^{m_3}_{n_3-1}\sqrt{n_3(n_1+1)}\right)\\
\hspace{-5mm}J_3{}^{m_0m_1m_2m_3}_{n_0n_1n_2n_3} = i\delta^{-m_0+m_1+m_2+m_3}_{-n_0+n_1+n_2+n_3}\delta^{m_3}_{n_3}\delta^{m_0}_{n_0}\left(\delta^{m_1}_{n_1+1}\sqrt{(n_1+1)n_2}-\delta^{m_1}_{n_1-1}\sqrt{n_1(n_2+1)}\right)\label{generators_for_Hermite_MHS2}
\end{align}
\subsection{Linear solutions and helicity}
In this subsection, we focus on the particular number of dimensions, i.e.\  $d=4$, and find the helicities of the plane wave solutions for the massless MHSYM model. We can use the expansion (\ref{HermiteExpansionPotential}) and insert it into linearised EoM obtained from (\ref{ymeom}). The component fields in the expansion satisfy
\begin{equation}
\label{LinearHermiteEoM}
\Box h_a^{n_0n_1n_2n_3}(x) - \partial_a \partial^b h_b^{n_0n_1n_2n_3}(x)= 0
\end{equation}
and they enjoy a linearised gauge symmetry of the form
\begin{equation}
\label{linearHermiteSymmetry}
\delta_\varepsilon h_a^{n_0n_1n_2n_3}(x) = \partial_a \varepsilon^{n_0n_1n_2n_3}(x)\,.
\end{equation}
To find out about the helicity of the field, we can write down a plane wave solution to the EoM (\ref{LinearHermiteEoM}), use the freedom available through (\ref{linearHermiteSymmetry}) to fix the gauge and choose a direction of propagation (conventionally, we choose the $z$-direction). Such a solution is given by
\begin{equation}
h^{a\,n_0n_1n_2n_3}_{\pm}(x) = \epsilon_{(\pm)}^a \,p^{n_0n_1n_2n_3}e^{ikx}\label{linear_MHS_field}\,,
\end{equation}
where $k^2=0$, meaning that the field is massless, and where
$
\epsilon_{(\pm)}^a = \frac{1}{\sqrt{2}}\begin{pmatrix}
0&
1&
\pm i&
0
\end{pmatrix}
$,
and $p^{n_0n_1n_2n_3}$ is an a priori unconstrained polarisation factor in the infinite-dimensional unitary representation of the Lorentz group. 

The helicity of a plane wave can be calculated as the eigenvalue of the rotation generator around the propagation axis. As we have followed the convention to choose the $z$-axis as the axis of propagation, we want to find the eigenvalue of the rotation operator $J_3$. When acting on (\ref{linear_MHS_field}), which is in a mixed representation, each generator will have two parts; one belonging to the finite-dimensional representation (indices $a,b$), and one belonging to the infinite-dimensional representation (indices $m_0,...,m_3$ and $n_0,...,n_3$ in case of the Hermite expansion in $d=4$), i.e.
\begin{equation}\label{J3split}
\left(J_3\right)^{m_0m_1m_2m_3}_{n_0n_1n_2n_3}{}^a{}_b = \left(J_3\right)^{m_0m_1m_2m_3}_{n_0n_1n_2n_3}\delta^a_b + \delta^{m_0m_1m_2m_3}_{n_0n_1n_2n_3}\left(J_3\right)^a{}_b\,,
\end{equation}
where $\left(J_3\right)^a{}_b$ is in the fundamental (vector) representation of the Lorentz group, given explicitly by
\begin{equation}
J_3=
\begin{pmatrix}
0 & 0 & 0 & 0 \\
0& 0 & -i & 0 \\[3pt]
0& i &  0 & 0\\[3pt]
0& 0 & 0 & 0\\	
\end{pmatrix}\,.
\end{equation}
We have found the eigenvectors of $\left(J_3\right)^{m_0m_1m_2m_3}_{n_0n_1n_2n_3}$ in \cite{Cvitan:2024xmn}. They are given by
\begin{equation}
p^{n_0n_1n_2n_3} = d^{n_0n_3}C^{n_1n_2}_{(r,\lambda)}\,,
\end{equation}
with the coefficients $d^{n_0n_3}$ arbitrary and $C^{n_1n_2}_{(r,\lambda)}$ given by
\begin{align}
C_{(r,\lambda)}^{n_1, n_2} 
=\delta^{n_1+n_2}_{r} i^{k-n_1-1} d^{r/2}{}_{k-r/2,n_1-r/2}\left( \frac\pi2 \right)\, , 
\label{eigenvectorJ3}
\end{align}
where $d^l{}_{m,n}(\beta)$ are the Wigner $d$-functions, $\lambda = (2k - r)$ with $r=n_1+n_2$ and $k$ having possible values $k =0,1,...,r\,.$
These coefficients, when contracted with the basis vectors, produce the expected azimuthal dependence in the auxiliary space, $C_{(r,\lambda)}^{n_1, n_2} f_{n_1}(u_x)f_{n_2}(u_y) \sim e^{i \lambda u_\phi}$, where summation convention for $n_1$ and $n_2$ is assumed.
It is then straightforward to see that 
\begin{equation}\label{totalhelicity}
\left(J_3\right)^{m_0m_1m_2m_3}_{n_0n_1n_2n_3}{}^a{}_b \epsilon_{(\pm)}^b \,p^{n_0n_1n_2n_3} = (\pm 1 + \lambda) \epsilon_{(\pm)}^a \,p^{m_0m_1m_2m_3}\,,
\end{equation}
where the polarisation coefficients of definite $(r,\lambda)$ are
\begin{equation}\label{prlambda}
p^{n_0n_1n_2n_3} = d^{n_0n_3}C^{n_1n_2}_{(r,\lambda)}\,.
\end{equation}
All together, this means that the plane waves (\ref{linear_MHS_field}) with $p^{n_0n_1n_2n_3}$ given by (\ref{prlambda}) for some given $r=0,1,\ldots$ and $\lambda=-r, -r+2,,\ldots,r-2,r$ can carry helicity (\ref{totalhelicity}) in range $-r-1,-r+1,\ldots,r-1,r+1$.
A single value of helicity can appear in infinitely many polarisation factors (\ref{prlambda}), e.g.\ helicity $3$ can appear for each even value of $r \geq 2$. We can also see that helicity is not a Lorentz invariant quantity for solutions such as (\ref{linear_MHS_field}). For example, if we boost the solution in the $x$ direction, $n_0$ and $n_1$ indices will get mixed, and the final expression will no longer have a well-defined helicity, i.e., it will be a superposition of terms with various values of $\lambda$. This is a characteristic of continuous spin particles, as emphasised in \cite{Schuster:2013pxj}.

Another approach to learn about supported helicities already noted in \cite{Cvitan:2021qvm} involves using spherical harmonics in the auxiliary space, which are diagonal in the rotation generator around the axis of propagation. The axis of propagation $\hat{\mathbf{k}}$ in spacetime defines the preferred axis $\hat{\mathbf{u}}_{\hat{\mathbf{k}}} \equiv \hat{\mathbf{k}}$ in the auxiliary space, as well as corresponding spherical coordinates $u_{r}$, $u_{\hat{\mathbf{k}},\theta}$ and $u_{\hat{\mathbf{k}},\phi}$ (where $\hat{\mathbf{u}}_{\hat{\mathbf{k}}}$ is tangent to the direction where $u_{\hat{\mathbf{k}},\theta}=0$). 
\begin{equation}
g_{r_0 n l m}(u,\hat{\mathbf{k}}) = f_{n_0}(u_0)\, F_n(u_{r})\, Y_l^m(u_{\hat{\mathbf{k}},\theta},u_{\hat{\mathbf{k}},\phi}),
\end{equation}
where $F_n$ are Laguerre functions, $Y_l^m$ are spherical harmonics, and $n_0 = 0, 1, 2, \ldots$, $n = 0, 1, 2, \ldots$, 
$l = 0, 1, 2, \ldots$,
, $m = -l, -l+1, \ldots, l$. The plane wave solutions for the MHS potential can then be expanded as
\begin{equation}
\mbox{\boldmath$\epsilon$}_{\sigma r_0 n l m}(\mathbf{k})\, e^{i k \cdot x}\, g_{r_0 n l m}(u,\hat{\mathbf{k}}) \quad,\quad 
\mathbf{k} \cdot \mbox{\boldmath$\epsilon$}_{\sigma r_0 n l m}(\mathbf{k}) = 0 \quad,\quad  k^2 = 0,
\end{equation}
with $\sigma = \pm1$. Helicity is $\sigma + m$, showing an infinite number of degrees of freedom for each helicity value.

\section{Wigner's classification}
\label{Wignerology}
The field we worked with above is massless, and to classify the possible excitations further, we need to find the value of the quartic Casimir operator of the Poincar\'e group. Here we would like to review the basics of Wigner's method for classifying elementary particles and highlight the relationship to the plane wave solutions of the linear equations of motion. This exposition closely follows the first volume of Weinberg's Quantum Theory of Fields \cite{Weinberg:1995mt} while additional details can be found in \cite{Loebbert:2008zz,Duncan:2012aja,Bekaert:2006py}.  Wigner's classification of elementary particles \cite{Wigner:1939cj} is a classification of a spacetime isometry group (in our case, the Poincaré group) represented on the space of one particle states. 

In $d=4$, the Poincaré group has a quadratic and a quartic Casimir operator
\begin{align}
C_2 = - P^\mu P_\mu=-P^2,\quad C_4 = W^\mu W_\mu = W^2\,,
\end{align}
where $P^\mu$ are the translation generators, the Pauli-Lubanski vector  $W^\mu$ is defined as
\begin{equation}
W_\rho = \frac{1}{2}\epsilon_{\mu\nu\rho\kappa}M^{\mu\nu}P^\kappa
\end{equation}
and the Lorentz generators are denoted by $M^{\mu\nu}$.
It is convenient to label the single-particle states with the eigenvalues of the Casimir operator. Further, since momentum operators form an abelian subgroup, we work with their eigenvectors and label them as
$ \ket{p^2, w^2, p^\mu,\sigma}$ 
such that 
\begin{equation}
P^2\ket{p^2, w^2, p^\mu,\sigma} = p^2\ket{p^2, w^2, p^\mu,\sigma},\quad W^2 \ket{p^2, w^2, p^\mu,\sigma} = w^2 \ket{p^2, w^2, p^\mu,\sigma}\label{Particle_Casimirs}
\end{equation} and
\begin{equation}
P^\mu \ket{p^2, w^2, p^\mu,\sigma} = p^\mu \ket{p^2, w^2, p^\mu,\sigma}
\end{equation}with $\sigma$ labelling other degrees of freedom which are to be determined.
\begin{table}[ht!]\centering
	\begin{tabular}{c|c|c|c}\hline
		$C_2 = - p^2$ & Standard momentum $k^\mu$ & Little group & Example \\
		\hline$p^2=-m^2$ & $(\pm m,0,0,0)$ & $SO(3)$ & $W^{\pm}$ and $Z$ bosons\\
		$p^2 = 0$ & $(\pm \omega,0,0,\omega)$ & $ISO(2)$& photons\\
		$p^2=n^2$ & $(0,0,0,n)$ & $SO(2,1)^\uparrow$& tachyons \\
		$p^2=0$ & $(0,0,0,0)$ & $SO(3,1)^\uparrow$& vacuum\\\hline
	\end{tabular}
	\caption{Representations of the Poincaré group}
	\label{Positive energy representations of the Poincaré group}
\end{table}

\noindent While translations act on the basis vectors as
\begin{equation}
U(\Delta x)\ket{p^2, w^2, p^\mu,\sigma} = e^{-ip\cdot\Delta x}\ket{p^2, w^2,  p^\mu,\sigma}\,,
\end{equation}
it can be shown that homogeneous Lorentz transformations act as
\begin{equation}
U(\Lambda)\ket{p^2, w^2,  p^\mu,\sigma} = \mathcal{N} \sum_{\sigma'} \mathcal{D}_{\sigma'\sigma}(W(\Lambda,p))\ket{p^2, w^2, \Lambda^\mu{}_\nu p^\nu,\sigma'}\,,
\end{equation}
where $\mathcal{N}=\mathcal{N}(p^0)$ is a normalisation factor and $W(\Lambda,p) = S^{-1}(\Lambda p)\Lambda S(p)$ is an element of the little group for a particular standard momentum with $S(p)$ defined to satisfy $p^\mu = S^\mu{}_\nu k^\nu$. Matrices $\mathcal{D}_{\sigma'\sigma}(W)$ furnish an irreducible representation of the little group, and their construction is sufficient to properly characterise the one-particle state. Within a choice of standard momentum, the quartic Casimir of the Poincaré group is equal to the Casimir operator of the little group. We will be especially interested in the massless case, so we will focus in more detail on the group $ISO(2)$. 

If we choose the momentum to be standard $p^\mu = k^\mu = (\omega, 0,0,\omega)$, we can explicitly find the components of the Pauli-Lubanski vector, which are the generators of the little group for the case of massless particles
\begin{equation}
W^\mu = \omega (J_3, J_1-K_2, J_2+K_1,J_3)\,.
\end{equation}
Where $J_i$ are generators of rotations and $K_i$ generators of boosts. It is convenient to name the generators
\begin{equation}\label{define_A_B}
A =\omega( J_1-K_2),\quad B = \omega(J_2+K_1)\,.
\end{equation}
It is easy to check that $A,B$ and $J_3$ span the Lie Algebra $\mathfrak{iso}(2)$
\begin{align}
[A, B] &= 0, \quad [J_3, A] = i B, \quad [J_3, B] = -iA\,.
\end{align}
The quartic Casimir in this choice of standard momentum is then given by
\begin{align}
W_\mu W^\mu\equiv W^2 =&\omega^2 (J_1-K_2)^2 + \omega^2(J_2+K_1)^2
\\
=& A^2 + B^2 = w^2\label{quartic_massless_onshell}\,.
\end{align}

The faithful irreducible unitary representations of the little group $ISO(2)$, which have a non-vanishing value of the Casimir operator $W^2$, are necessarily infinite dimensional \cite{Tung:1985na}. If written in a basis diagonal in the rotation operator around the standard momentum, it can be seen that each irreducible representation contains an infinite tower of helicity states mixing under Lorentz transformations. For that reason, such representations are usually named ``infinite-spin''. A different basis choice is possible, which motivates a different name --- ``continuous spins''. This class of representations was originally considered by Wigner to be unsuitable for physical use since the infinite tower of helicities would have to correspond to an infinite heat capacity. However, in recent years, there has been a revived interest in this class of particles and in analysing their kinematic and dynamical aspects (\cite{Schuster:2013pta,Schuster:2013pxj,Schuster:2013vpr,Schuster:2014hca,Bekaert:2017khg,Rivelles:2016rwo,Schuster:2023xqa,Schuster:2023jgc,Schuster:2024wjc}).

There is also a possibility of a non-faithful representation of the little group $ISO(2)$ where the operators $A,B$ act trivially. In this case, $W^2$ gives a vanishing value, and the little group becomes isomorphic to $SO(2)$. The representations are one-dimensional, with the only non-trivial operator being the rotation around the standard momentum. The eigenvectors of this rotation are the ordinary helicity states (due to eigenvalues of $A$ and $B$ being zero, helicity is now Lorentz invariant) describing particles corresponding to massless fields of a fixed spin such as the Maxwell field, linear Einstein gravity, higher spin fields of the Fronsdal type, etc.

There is an important relationship between the matrices $\mathcal{D}_{\sigma'\sigma}(W)$, which, as we have seen, act on the one-particle states, and the Lorentz transformation matrices we use to express quantum field components in different inertial frames.  From the creation and annihilation operators, we can build a quantum field\begin{equation}
h^{r}(x) = \sum_{\sigma} \int \frac{d^3 \textbf{p}}{(2\pi)^{3/2}\sqrt{2\omega_\textbf{p}}}\left(u^r(\textbf{p},\sigma)a(\textbf{p},\sigma)e^{-ipx} + v^r(\textbf{p},\sigma)a^\dagger(\textbf{p},\sigma)e^{ipx}\right)\,,
\end{equation} where $r$ stands for any set of Lorentz indices (e.g.\ for the MHS model, this includes both $a$ and $\mu$ indices in case of the Taylor expansion and $a$ and $n$ indices in case of the Hermite expansion).
Under a Lorentz transformation, it transforms as
\begin{equation}
U(\Lambda)h^r(x)U^{-1}(\Lambda) = \sum_{s} D(\Lambda^{-1})^{rs} h^s(\Lambda x)\,,
\end{equation}
where $D(\Lambda)$ is a representation of the Lorentz group (finite or infinite-dimensional) on the space of fields. This equation can be seen as a compatibility condition \cite{Duncan:2012aja,Weinberg:1995mt,Tung:1985na} between the infinite-dimensional unitary Fock-space representation on the one-particle states and the representation of the homogeneous Lorentz group on the space of fields. In a straightforward manner, it can be brought down to compatibility equations for the polarisation functions in the standard momentum $\mathbf{k}$
on the level of the Lie algebra of the little group
\begin{align}
\sum_{\sigma'}u^r(\mathbf{k},\sigma')\mathcal{J}_{\sigma'\sigma}=\sum_s J^{rs} u^s(\mathbf{k},\sigma)\\
\sum_{\sigma'}v^r(\mathbf{k},\sigma')\mathcal{J}^*_{\sigma'\sigma} =-\sum_s J^{rs}v^s(\mathbf{k},\sigma)\,,
\end{align}
which follows from the expansion $\mathcal{D}_{\sigma'\sigma} \approx \delta_{\sigma\sigma'} + i\theta \mathcal{J}_{\sigma\sigma'}$, and $L^{rs} \approx \delta^{rs} + i\theta J^{rs}$. 

The polarisation functions in the standard momentum thus carry the representation of the little group, as do the one-particle states (\ref{Particle_Casimirs}), and also contain information about the quartic Casimir operator. By classifying polarisation functions of a certain field, through solving the eigensystem
\begin{align}
\sum_s D\left(W^2\right)^{rs} u^s(\mathbf{k},\sigma) = w^2 u^r(\mathbf{k},\sigma)\label{eigensystem_from_qft}\,,
\end{align}
where $w^2$ on the right-hand side is constant due to (\ref{Particle_Casimirs}), we can learn about supported particle types, i.e.\ values of the Casimir operator $W^2$, associated with that particular field.

\section{The quartic Casimir in the Hermite expansion}\label{sectionV}
Our first goal is to build an explicit expression for (\ref{quartic_massless_onshell}) for the case of the plane wave solution (\ref{linear_MHS_field}) and then attempt to find possible eigenvalues. For the sake of the clarity of argument, we will demonstrate the procedure on the Maxwell field before following these steps for the MHSYM component field.

In the case of electrodynamics, a plane-wave solution of the equations 
\begin{equation}
\Box A^\mu -\partial^\mu \partial \cdot A = 0\,,
\end{equation} with momentum oriented in the $z$ direction $k^\mu = (\omega, 0, 0, \omega)$ is given by $A^{\mu}(x) =\epsilon_{\pm}^\mu e^{ikx}$ where $\epsilon_{\pm}^\mu = (0,1,\pm i, 0)$.
Since $A^\mu(x)$ is a Lorentz vector, we can use the vector representation of the Lorentz generators $\left(M^{\mu\nu}\right)^\rho{}_{\kappa} = i\left(\eta^{\mu\rho}\delta^{\nu}_\kappa - \eta^{\nu\rho}\delta^{\mu}_\kappa \right)\,$.
The quartic Casimir element (\ref{quartic_massless_onshell}) is then explicitly given by
\begin{equation}
W^2 = \omega^2
\begin{pmatrix}
-2 & 0 & 0 & 2 \\
0 & 0 & 0 & 0 \\
0 & 0 & 0 & 0 \\
-2 & 0 & 0 & 2 \\
\end{pmatrix}\,,
\end{equation}
so it is readily visible through applying (\ref{eigensystem_from_qft}) that there is a single eigenvalue of the quartic Casimir, and it is vanishing $\left(W^2\right)^\mu{}_\nu \, A^\nu = 0\,.$
\subsection{Quartic Casimir of the on-shell MHS field}
As stated, the MHS field is in a mixed representation of the Lorentz group - a direct product of the finite-dimensional vector representation and the infinite-dimensional unitary representation. As in (\ref{J3split}), the generators will be a direct sum of two parts: one belonging to the finite-dimensional representation and one belonging to the infinite-dimensional representation, e.g.
\begin{equation}
\left(J_1\right)^{m_0m_1m_2m_3}_{n_0n_1n_2n_3}{}^a{}_b = \left(J_1\right)^{m_0m_1m_2m_3}_{n_0n_1n_2n_3}\delta^a_b + \delta^{m_0m_1m_2m_3}_{n_0n_1n_2n_3}\left(J_1\right)^a{}_b\,.
\end{equation}
Using capital $\mathbf{N}$ instead of the tuple $\{n_0n_1n_2n_3\}$
\begin{equation}
\left(J_1\right)^{\textbf{M}}_{\textbf{N}}{}^a{}_b = \left(J_1\right)^{\textbf{M}}_{\textbf{N}}\delta^a_b + \delta^{\textbf{M}}_{\textbf{N}}\left(J_1\right)^a{}_b\,.
\end{equation}
The Casimir element (\ref{quartic_massless_onshell}) is then given by (repeated indices summed over)
\begin{align*}\label{Full_MHS_Casimir}
\left(W^2\right)^\textbf{M}_\textbf{N}{}^a{}_c =& \omega^2\Big(J_1^a{}_b J_1^b{}_c +J_2^a{}_b J_2^b{}_c+K_1^a{}_b K_1^b{}_c+K_2^a{}_b K_2^b{}_c-2J_1^a{}_bK_2^b{}_c +2K_1^a{}_b J_2^b{}_c\Big)\delta^\textbf{M}_\textbf{N} \\
+& \omega^2\Big(J_1{}^\textbf{M}_\textbf{R} J_1{}^\textbf{R}_\textbf{N} +J_2{}^\textbf{M}_\textbf{R} J_2{}^\textbf{R}_\textbf{N} + K_1{}^\textbf{M}_\textbf{R} K_1{}^\textbf{R}_\textbf{N} + K_2{}^\textbf{M}_\textbf{R} K_2{}^\textbf{R}_\textbf{N}- 2 J_1{}^\textbf{M}_\textbf{R} K_2{}^\textbf{R}_\textbf{N} + 2 K_1{}^\textbf{M}_\textbf{R} J_2{}^\textbf{R}_\textbf{N}\Big)\delta^a_c\\
+&\omega^2\Big( 2J_1^a{}_cJ_1{}^\textbf{M}_\textbf{N} + 2J_2^a{}_cJ_2{}^\textbf{M}_\textbf{N} + 2K_1^a{}_cK_1{}^\textbf{M}_\textbf{N} +  2K_2^a{}_cK_2{}^\textbf{M}_\textbf{N}\\
-& 2K_2^a{}_cJ_1{}^\textbf{M}_\textbf{N} - 2K_2{}^\textbf{M}_\textbf{N}J_1^a{}_c + 2K_1^a{}_cJ_2{}^\textbf{M}_\textbf{N} + 2K_1{}^\textbf{M}_\textbf{N}J_2^a{}_c\Big)\,.\numberthis
\end{align*}
The first bracket contains the finite-dimensional vector representation of the $W^2$, which is multiplied by $\delta^\textbf{M}_\textbf{N}$. As in the case of a finite-dimensional massless vector field, this will give zero when acting on the polarisation vector $\epsilon^a$ found in (\ref{linear_MHS_field}).

The mixed contributions to the Casimir can be rewritten as
\begin{align*}
(W^2_{mixed})^\textbf{M}_\textbf{N}{}^a{}_c 
=& 2A^a{}_cA^{\textbf{M}}_{\textbf{N}} + 2B^a{}_cB^{\textbf{M}}_{\textbf{N}}\,,
\end{align*}
where $A$ and $B$ were defined in (\ref{define_A_B}). When $A^{a}{}_c$ or $B^{a}{}_c$ act on the polarisation vector $\epsilon^a$, the result will be proportional to the momentum, e.g.\ for $A^{a}{}_c$
\begin{equation}
i\left(
\begin{array}{cccc}
0 & 0 & -1 & 0 \\
0 & 0 & 0 & 0 \\
-1 & 0 & 0 & 1 \\
0 & 0 & -1 & 0 \\
\end{array}
\right) \frac{1}{\sqrt{2}}\begin{pmatrix}
0\\
1\\
\pm i\\
0
\end{pmatrix} =  \frac{\pm 1}{\sqrt{2}} \begin{pmatrix}
1\\
0\\
0\\
1
\end{pmatrix}\propto p^a\,,
\end{equation}
i.e.\ a pure gauge contribution in the finite-dimensional sector. 
A non-trivial eigenvalue of the quartic Casimir for the MHS field can thus come only from the second line of (\ref{Full_MHS_Casimir}), which contains the infinite-dimensional part
\begin{align}
\hspace{-0.5cm}\left(W^2_{inf}\right)^\textbf{M}_\textbf{N}{\delta}^a{}_c = \omega^2\delta^a_c (J_1{}^\textbf{M}_\textbf{R} J_1{}^\textbf{R}_\textbf{N} + J_2{}^\textbf{M}_\textbf{R} J_2{}^\textbf{R}_\textbf{N} +K_1{}^\textbf{M}_\textbf{R} K_1{}^\textbf{R}_\textbf{N} + K_2{}^\textbf{M}_\textbf{R} K_2{}^\textbf{R}_\textbf{N} - 2K_2{}^\textbf{M}_\textbf{R} J_1{}^\textbf{R}_\textbf{N} + 2J_2{}^\textbf{M}_\textbf{R} K_1{}^\textbf{R}_\textbf{N})\,.
\end{align}
We now use the explicit expressions for the infinite-dimensional generators (\ref{generators_for_Hermite_MHS1}-\ref{generators_for_Hermite_MHS2}) and arrive at the result written out without the use of the compact notation. 
\begin{align*}
\hspace{-0.5cm}\left(W^2_{inf}\right)^{m_0m_1m_2m_3}_{n_0n_1n_2n_3} =& \omega^2\delta_{-n_0 + n_1 + n_2 + n_3}^{-m_0 + m_1 + m_2 + m_3}\times\\
\Bigg(&2\delta^{m_0}_{n_0}\delta^{m_1}_{n_1}\delta^{m_2}_{n_2}\delta^{m_3}_{n_3}(1+n_0+n_3)(1+n_1+n_2)\\
-&\delta^{m_0}_{n_0}\delta^{m_1}_{n_1}\delta^{m_2}_{n_2+2}\delta^{m_3}_{n_3-2}\sqrt{(n_2+1)(n_2+2)(n_3-1)n_3}\\
-&\delta^{m_0}_{n_0}\delta^{m_1}_{n_1}\delta^{m_2}_{n_2-2}\delta^{m_3}_{n_3+2}\sqrt{(n_2-1)n_2(n_3+2)(n_3+1)}\\
-&\delta^{m_0}_{n_0}\delta^{m_1}_{n_1-2}\delta^{m_2}_{n_2}\delta^{m_3}_{n_3+2}\sqrt{(n_1-1)n_1(n_3+1)(n_3+2)} \\
-&\delta^{m_0}_{n_0}\delta^{m_1}_{n_1+2}\delta^{m_2}_{n_2}\delta^{m_3}_{n_3-2}\sqrt{(n_1+1)(n_1+2)(n_3-1)n_3}\\
-&\delta^{m_0}_{n_0+2}\delta^{m_1}_{n_1+2}\delta^{m_2}_{n_2}\delta^{m_3}_{n_3}\sqrt{(n_0+1)(n_0+2)(n_1+1)(n_1+2)} \\
- &\delta^{m_0}_{n_0-2}\delta^{m_1}_{n_1-2}\delta^{m_2}_{n_2}\delta^{m_3}_{n_3}\sqrt{(n_0-1)n_0(n_1-1)n_1}\\
-&\delta^{m_0}_{n_0+2}\delta^{m_1}_{n_1}\delta^{m_2}_{n_2+2}\delta^{m_3}_{n_3}\sqrt{(n_0+1)(n_0+2)(n_2+2)(n_2+1)} \\
- &\delta^{m_0}_{n_0-2}\delta^{m_1}_{n_1}\delta^{m_2}_{n_2-2}\delta^{m_3}_{n_3}\sqrt{(n_0-1)n_0(n_2-1)n_2}\\
+&2\delta^{m_0}_{n_0+1}\delta^{m_1}_{n_1}\delta^{m_2}_{n_2 + 2}\delta^{m_3}_{n_3 - 1}\sqrt{(n_0+1)(n_2+1)(n_2+2)n_3} \\
-&2\delta^{m_0}_{n_0+1}\delta^{m_1}_{n_1}\delta^{m_2}_{n_2}\delta^{m_3}_{n_3 + 1}\left(\sqrt{(n_0+1)n_2n_2(n_3+1)}+\right.\\
&\qquad\qquad\qquad\qquad \left. +\sqrt{(n_0+1)(n_1+1)(n_1+1)(n_3+1)}\right)\\
-&2\delta^{m_0}_{n_0-1}\delta^{m_1}_{n_1}\delta^{m_2}_{n_2}\delta^{m_3}_{n_3 - 1}\left(\sqrt{n_0(n_2+1)(n_2+1)n_3}+\sqrt{n_0n_1n_1n_3}\right)\\
+ &2\delta^{m_0}_{n_0-1}\delta^{m_1}_{n_1}\delta^{m_2}_{n_2 - 2}\delta^{m_3}_{n_3 + 1}\sqrt{n_0(n_2-1)n_2(n_3+1)}\\
+&2\delta^{m_0}_{n_0+1}\delta^{m_1}_{n_1+2}\delta^{m_2}_{n_2}\delta^{m_3}_{n_3 - 1}\sqrt{(n_0+1)(n_1+1)(n_1+2)n_3} \\
+&2\delta^{m_0}_{n_0-1}\delta^{m_1}_{n_1-2}\delta^{m_2}_{n_2}\delta^{m_3}_{n_3 + 1}\sqrt{n_0(n_1-1)n_1(n_3+1)}\Bigg)\numberthis\label{Casimir_full}\,.
\end{align*}
When acting on the field polarisation factors $p^{n_0n_1n_2n_3}$ in (\ref{linear_MHS_field}), we get
\begin{align*}\label{Casimir_on_polarisations}
\left(W^2_{inf}\right)&^{m_0m_1m_2m_3}_{n_0n_1n_2n_3} p^{n_0n_1n_2n_3}=\\
2&p^{m_0m_1m_2m_3}\left[(1+m_0+m_3)(1+m_1+m_2)\right]\\ 
-&p^{m_0m_1(m_2-2)(m_3+2)}\sqrt{(m_2-1)m_2(m_3+1)(m_3+2)}\\
-&p^{m_0m_1(m_2+2)(m_3-2)}\sqrt{(m_2+1)(m_2+2)m_3(m_3-1)}\\
-&p^{m_0(m_1+2)m_2(m_3-2)}\sqrt{(m_1+1)(m_1+2)(m_3-1)m_3}\\
-&p^{m_0(m_1-2)m_2(m_3+2)}\sqrt{(m_1-1)m_1(m_3+1)(m_3+2)}\\
-&p^{(m_0-2)(m_1-2)m_2m_3}\sqrt{(m_0-1)m_0(m_1-1)m_1}\\
-&p^{(m_0+2)(m_1+2)m_2m_3}\sqrt{(m_0+1)(m_0+2)(m_1+1)(m_1+2)}\\
-&p^{(m_0-2)m_1(m_2-2)m_3}\sqrt{(m_0-1)m_0m_2(m_2-1)}\\
-&p^{(m_0+2)m_1(m_2+2)m_3}\sqrt{(m_0+1)(m_0+2)(m_2+1)(m_2+2)}\\
+2&p^{(m_0-1)m_1(m_2-2)(m_3+1)}\sqrt{m_0(m_2-1)m_2(m_3+1)}\\
-2&p^{(m_0-1)m_1m_2(m_3-1)}\sqrt{m_0m_2m_2m_3)}\\
-2&p^{(m_0+1)m_1m_2(m_3+1)}\sqrt{(m_0+1)(m_2+1)(m_2+1)(m_3+1)}\\
+2&p^{(m_0+1)m_1(m_2+2)(m_3-1)}\sqrt{(m_0+1)(m_2+1)(m_2+2)m_3}\\
-2&p^{(m_0-1)m_1m_2(m_3-1)}\sqrt{m_0(m_1+1)(m_1+1)m_3}\\
+2&p^{(m_0-1)(m_1-2)m_2(m_3+1)}\sqrt{m_0(m_1-1)m_1(m_3+1)}\\
+2&p^{(m_0+1)(m_1+2)m_2(m_3-1)}\sqrt{(m_0+1)(m_1+1)(m_1+2)m_3}\\
-2&p^{(m_0+1)m_1m_2(m_3+1)}\sqrt{(m_0+1)m_1m_1(m_3+1)}\numberthis\,.
\end{align*}
\subsection{Casimir eigenvalue problem}
To learn about the particle spectrum of our theory, following (\ref{eigensystem_from_qft}), we should solve the eigensystem
\begin{equation}\label{eigensystem_casimir}
\left(W^2_{inf}\right)^{m_0m_1m_2m_3}_{n_0n_1n_2n_3} p^{n_0n_1n_2n_3} = w^2 p^{m_0m_1m_2m_3}\,.
\end{equation}
Even before explicitly trying to find eigenvectors and eigenvalues in (\ref{eigensystem_casimir}), we can conclude from (\ref{Casimir_on_polarisations}) that there will exist non-trivial states, i.e.\ the expression (\ref{Casimir_on_polarisations}) shows that a polarisation factor $p^{n_0n_1n_2n_3}$ used in (\ref{linear_MHS_field}) will in general not give a vanishing eigenvalue, through which we can confirm that the MHS formalism supports a description of the infinite spin particles.

One way of tackling the eigenvalue problem is by computer-assisted iterative solving, which could give us a hint for an appropriate ansatz. It can be seen by the structure of the Casimir that any eigenvector should necessarily have an infinite number of components 
(terms such as $\delta^{m_0}_{n_0-1}\delta^{m_1}_{n_1}\delta^{m_2}_{n_2}\delta^{m_3}_{n_3 - 1}$ will always simultaneously raise the values of the 0th and 3rd index; thus a closed solution cannot have a finite number of terms), 
so we could only hope for hints coming from a truncated calculation. The Casimir operator can be rewritten in a basis of eigenvectors of $J_3$, which we found in \cite{Cvitan:2024xmn}, but the complexity of the problem remains. So far, we are left with finding educated guesses, and one of them comes from the ``massless limit'' of eigenvectors of $\vec{J}^2$.
\subsubsection{Massless limit of massive states}
In the case of massive particles, the little group is $SO(3)$, and the Casimir operator is simply $\vec{J}^2$. An appropriately performed Inönü-Wigner contraction of a representation of $SO(3)$ can give us a representation of $ISO(2)$, which is the little group in the case of massless particles. The limiting procedure entails the limits $m\to0,v\to 1$ while keeping fixed $m\gamma = \frac{m}{\sqrt{1-v^2}} = \omega$. In the representation by Hermite functions, we have
\begin{align*}
(\vec{J})^2\,{}^{m_0m_1m_2m_3}_{n_0n_1n_2n_3} =& \delta^{-m_0+m_1+m_2+m_3}_{-n_0+n_1+n_2+n_3}\delta^{m_0}_{n_0}\times\\
\Big(2&\delta^{m_1}_{n_1}\delta^{m_2}_{n_2}\delta^{m_3}_{n_3}(n_1+n_2+n_3 + n_1n_2 + n_2n_3 + n_3n_1)\\ 
-&\delta^{m_1}_{n_1}\delta^{m_2}_{n_2-2}\delta^{m_3}_{n_3+2}\sqrt{(n_2-1)n_2(n_3+1)(n_3+2)}\\ 
-&\delta^{m_1}_{n_1}\delta^{m_2}_{n_2+2}\delta^{m_3}_{n_3-2}\sqrt{(n_2+1)(n_2+2)(n_3-1)n_3}\\
-&\delta^{m_1}_{n_1+2}\delta^{m_2}_{n_2}\delta^{m_3}_{n_3-2}\sqrt{(n_3-1)n_3(n_1+1)(n_1+2)}\\ 
-&\delta^{m_1}_{n_1-2}\delta^{m_2}_{n_2}\delta^{m_3}_{n_3+2}\sqrt{(n_3+1)(n_3+2)(n_1-1)n_1}\\
-&\delta^{m_1}_{n_1-2}\delta^{m_2}_{n_2+2}\delta^{m_3}_{n_3}\sqrt{(n_1-1)n_1(n_2+1)(n_2+2)}\\ 
-&\delta^{m_1}_{n_1+2}\delta^{m_2}_{n_2-2}\delta^{m_3}_{n_3}\sqrt{(n_1+1)(n_1+2)(n_2-1)n_2}\,\Big)\,.\label{so(3)Casimir}\numberthis
\end{align*}

The simplest simultaneous eigenvector of $\vec{J}^2$ (\ref{so(3)Casimir}) and $J_3$ (\ref{generators_for_Hermite_MHS2}) in the representation over Hermite functions is 
\begin{equation}
\Phi_{n=0,s=0,\lambda=0}(u) =\delta^{n_0}_0 \delta^{n_1}_0\delta^{n_2}_0\delta^{n_3}_0f_{n_0n_1n_2n_3}(u)\label{simplest_eigenvector_of_JJ}\,,
\end{equation}
where $n=n_1+n_2+n_3$ and $s$ corresponds to the eigenvalue of $\vec{J}^2=s(s+1)$ while $\lambda$ is an eigenvalue of $J_3$. We can boost (\ref{simplest_eigenvector_of_JJ}) with velocity $v$ in the $z$ direction to prepare it for the massless limit.
The transformation matrices for a finite boost are (see \cite{Cvitan:2024xmn} and  (\ref{representation_matrices_sign_change}))
\begin{align}
{L}_{n_0n_1n_2n_3}^{m_0m_1m_2m_3}&(v) = (-1)^{m_0+m_3}\sqrt{\frac{m_3!n_0!}{n_3!m_0!}}\delta_{-n_0+n_1+n_2+n_3}^{-m_0 + m_1+ m_2 + m_3}\delta^{m_1}_{n_1}\delta^{m_2}_{n_2}\times\nonumber\\
&\sum_{j=0}^{m_0}\binom{m_0}{j}\binom{n_3}{m_3-j}(-1)^{j}\sqrt{1-v^2}^{m_3+m_0+1-2j}v^{2j-m_3+n_3} \label{DmatrixBoostZ}\,.
\end{align}
Boosting the chosen eigenvector we get
\begin{align}
\sum_{n_0,n_1,n_2,n_3=0}^\infty L_{n_0n_1n_2n_3}^{m_0m_1m_2m_3}(v)\cdot \delta^{n_0}_0 \delta^{n_1}_0\delta^{n_2}_0\delta^{n_3}_0=(-1)^{m_0}\delta^{m_1}_0\delta^{m_2}_0\delta^{m_0, m_3}\sqrt{1-v^2}(-v)^{m_3}\,.
\end{align}
Since $m\gamma \to \omega$ while $v\to 1$, we can divide this result by $m$ to obtain a useful result in the limiting procedure
\begin{equation}
p^{n_0n_1n_2n_3} \to \delta^{m_1}_0\delta^{m_2}_0\delta^{m_0,m_3}\,.\label{massless_limit_of_0}
\end{equation}
A more general case could be an eigenvector of $J_3$ and $J^2$ of the form
\begin{equation}\label{general_eigenvector_J3J2}
\Phi_{n=n,s=n,\lambda=n}(u) = z^{(n_0)}C^{n_1n_2}_{(n,n)} \delta^{n_3}_0 f_{n_0n_1n_2n_3}(u)\,.
\end{equation}
In the sector $n_1+n_2+n_3 = n$ where $s=n$ and $\lambda = n$, there is only one such vector, and the factor $C^{n_1n_2}_{(n,n)}$ is given in (\ref{eigenvectorJ3}). The index $n_3$ has to be equal to $0$, while $n_0$ is arbitrary (or fixed by a choice of $N=-n_0+n$), meaning that the factor $z^{(n_0)}$ must have the form
\begin{equation}
z^{(n_0)} = const. \cdot \delta^{n_0}_{n-N}\,.
\end{equation}
We boost (\ref{general_eigenvector_J3J2}) in the $z$ direction to prepare it for the massless limit
\begin{align}
	L_{n_0n_1n_2n_3}^{m_0m_1m_2m_3}(v)z^{(n_0)}C^{n_1n_2}_{n,n} \delta^{n_3}_0 =& (-1)^{m_0}\sqrt{\frac{m_0!}{m_3!(m_0-m_3)!}}\nonumber\\
	&z^{m_0-m_3}C^{m_1m_2}_{(m_1+m_2,m_1+m_2)}\sqrt{1-v^2}^{m_0-m_3+1}(-v)^{m_3}\label{highest_weight_boosted}\,.
\end{align}
\subsubsection{Eigenvector candidates}
The result in (\ref{highest_weight_boosted}) motivates an ansatz of the form
\begin{align}
p^{n_0n_1n_2n_3} = \delta^{n_0,n_3}C^{n_1n_2}_{r,r}c^{(n_3)}\,.
\end{align}
Upon inserting into (\ref{eigensystem_casimir}) we obtain two independent equations for $c^{(m_3)}$
\begin{align}
c^{(m_3)}(1+2m_3)-c^{(m_3-1)}m_3-c^{(m_3+1)}(m_3+1) = -\frac{w^2}{2(1+r)} c^{(m_3)}\,,
\end{align}
\begin{equation}
c^{(m_3+2)}+c^{(m_3)}-2c^{(m_3+1)}=0\,.
\end{equation}
The solution is given by $w^2 = 0$ and
\begin{equation}
c^{(m_3)}= c^{(0)}\,.
\end{equation}
This gives the polarisation factor
\begin{align}\label{eigenvector_casimir_2}
p^{n_0n_1n_2n_3} = \delta^{n_0,n_3}C^{n_1n_2}_{(r,r)}\,,
\end{align}
which is simultaneously an eigenvector of $J_3$ with helicity $\lambda = r$.
The norm of this solution is not finite, and we can explicitly see that in the sum over Hermite functions in the auxiliary space, the eigenvector will contain a delta function. 
As an example with $r=0$, from the completeness identity of Hermite functions, we find
\begin{equation}
\sum_{n_0,n_1,n_2,n_3=0}^\infty\delta^{n_0,n_3}\delta^{n_1}_0\delta^{n_2}_0f_{n_0}(u_0)f_{n_1}(u_1)f_{n_2}(u_2)f_{n_3}(u_3) = \delta^{(2)}(u_0 - u_3)e^{-\frac{u_1^2+u_2^2}{2}}\,.
\end{equation}

Through this approach, we were able to obtain a solution to the equation (\ref{eigensystem_casimir}) with a vanishing eigenvalue of the quartic Casimir $W^2$ and with an arbitrary integer helicity. This would, by definition, correspond to an ordinary higher-spin massless field. Observe that this solution is not square integrable, which means that if it is present in the physical spectrum, it must belong to a continuous spectrum. To see this, a different approach will be used for the complete characterisation of the particle spectrum, which we show in the next section.
\section{The quartic Casimir for an on-shell master field}
\label{sectionVI}Consider the linearised equations of motion one obtains if the integration over the auxiliary space is not performed prior to extremising the action
\begin{equation}
\Box h_a(x,u) - \partial_a \partial^b h_b(x,u) = 0\,.
\end{equation}
As in the previous section, we can fix the gauge to $\partial^a h_a(x,u) = 0$, and consider solutions representing plane waves directed along the $z$-axis
\begin{equation}
h_a(x,u) = \epsilon_a \Phi(u) e^{ikx}\,,\label{master_space_linear_solution}
\end{equation}
where $k^a = (\omega,0,0,\omega)$ and $\epsilon_a = \frac{1}{\sqrt{2}}(0,1,\pm i, 0)$.
Now, let's consider an active Lorentz transformation following the transformation properties (\ref{LorentzActiveTransformation})
\begin{equation}
h'_a(x^c,u_d) = \Lambda_a{}^b h_b((\Lambda^{-1}x)^c,(u\cdot\Lambda)_d)\,.
\end{equation}
If we expand the Lorentz transformation matrix up to the first-order $$\Lambda^a{}_b \approx \delta^a_b + i \psi G^a{}_b\,,$$
with $\psi$ the expansion parameter
(the parameter can be an angle if $\Lambda$ is a rotation, or rapidity in case of boosts), 
then $(\Lambda^{-1})^a{}_b \approx \delta^a_b - i \psi G^a{}_b$ and since $(\Lambda^{-1})^a{}_b = \Lambda_b{}^a$ it is true that $G^a{}_b = - G_b{}^a$. Through a simple expansion, we get
\begin{align}
h'_a(x^c,u_d) 
\approx h_a(x,u) + i\psi (G_a{}^bh_b(x,u) + G_c{}^b x^c\partial_b^xh_a(x,u) +  G^b{}_cu_b\partial^c_u h_a(x,u))\,.
\end{align}
In the case of our solution (\ref{master_space_linear_solution}), the action of a generator of the Lorentz group, where $D(G)$ is a representation of the generator $G$, becomes
\begin{equation}
D(G)\cdot h_a(x,u) = \left(G_a{}^b\epsilon_b \Phi(u) + G_c{}^bx^cik_b\Phi(u) + G^b{}_c u_a\partial^c_u\Phi(u)\epsilon_b\right)e^{ikx}\label{generator_on_solution}\,.
\end{equation}
We would now like to examine the behaviour of (\ref{master_space_linear_solution}) under the action of the generators $A,B$ of the little group $\mathfrak{iso}(2)$ with the reference momentum $k^a = (\omega,0,0,\omega)$. If we are able to find eigenfunctions of the mentioned generators, they will be the on-shell basis for the representation of the little group. Since $A,B$ commute, their eigenfunctions will correspond to the plane-wave basis of $\mathfrak{iso}(2)$ seen in \cite{Tung:1985na}.

It is straightforward to find the explicit vector representations for the operators $A = J_1 - K_2$ and $B = J_2 + K_1$.
\begin{equation}
A = \omega
\begin{pmatrix}
0 & 0 & i & 0 \\
0&0 & 0 & 0 \\[3pt]
i&0 &  0 & i\\[3pt]
0&0 & -i & 0\\	
\end{pmatrix},\quad B = \omega
\begin{pmatrix}
0 & -i & 0 & 0 \\
-i&0 & 0 & -i \\[3pt]
0&0 &  0 & 0\\[3pt]
0&0 & i & 0\\	
\end{pmatrix}\,.
\end{equation}

We see from (\ref{generator_on_solution}) the three possible terms, of which only one will be non-trivial. Since $A^a{}_b k^b = B^a{}_b k^b = 0$, and  $A^a{}_b \epsilon^b \propto k^b, B^a{}_b \epsilon^b \propto k^b$, which is a pure gauge contribution, the only important term in the equation (\ref{generator_on_solution}) in the case of the generators $A$ and $B$ is the last one, of the form $G^b{}_c u_b\partial^c_u\Phi(u)$. We now explicitly state the differential equations for $A$ and $B$.
\begin{align}
A\cdot\Phi(u) =& u_{a} A^{a}{}_b  \partial_{u}^{b} \Phi(u)\\
=&i\omega (u_{t}-u_{z})\frac{\partial}{\partial u_{y}} + i \omega u_{y}\left(\frac{\partial}{\partial u_{t}} + \frac{\partial}{\partial u_{z}}\right) \Phi(u)\,.
\end{align}
In null-coordinates $u_+ = u_t + u_z, u_- = u_t - u_z$ it becomes somewhat simpler
\begin{align}
A\cdot\Phi(u) =i\omega \left[u_-\frac{\partial}{\partial u_y} + 2u_y\frac{\partial}{\partial u_+}\right]\Phi(u_+,u_-,u_x,u_y)\,.
\end{align}
The equation for $B$ is similarly
\begin{align}
B \cdot \Phi(u) = -i\omega \left[u_-\frac{\partial}{\partial u_x} + 2u_x \frac{\partial}{\partial u_+}\right]\Phi(u_+,u_-,u_x,u_y)\,.
\end{align} 
Similarly to (\ref{eigensystem_from_qft}) 
we want to find functions $\Phi(u)$ that satisfy the eigensystem
\begin{align}
A\cdot\Phi(u) &= \alpha\, \Phi(u)\\
B \cdot \Phi(u) &= \beta\, \Phi(u)\,.
\end{align}
The solutions to these equations for $A$ and $B$ separately are
\begin{align}
\Phi_A(u) = \exp\left({\frac{-i \alpha u_y}{\omega u_-}}\right)G_1\left(u_-, u_x, -(u_t)^2 + (u_y)^2 + (u_z)^2\right)\\
\Phi_B(u) = \exp\left({\frac{i \beta u_x}{\omega u_-}}\right)G_2\left(u_-, u_y, -(u_t)^2 + (u_x)^2 + (u_z)^2\right)\,,
\end{align}
where $G_1$ and $G_2$ are arbitrary functions of their respective variables. We can write down a simultaneous solution with $G_r$ an arbitrary function as
\begin{equation}\label{CasimirEigenfunction1}
\Phi_{\alpha \beta r}(u) = \exp\left({i\frac{\beta u_x - \alpha u_y}{\omega u_-}}\right)G_r\left(u_-,u_\mu u^\mu\right)\,,
\end{equation}
where $\alpha$ and $\beta$ stand for the eigenvalues of $A$ and $B$, and $r$ stands for any additional indices that may be used to discriminate between different solutions. The explicit representation for $W^2$ is
\begin{align}
&W^2 = A^2 + B^2 \\
=&-\omega^2\left(u_-^2 \left(\frac{\partial^2}{\partial u_x^2} + \frac{\partial^2}{\partial u_y^2}\right) + 4u_-\left(u_x\frac{\partial}{\partial u_x} +u_y\frac{\partial}{\partial u_y} + 1\right)\frac{\partial}{\partial u_+} + 4(u_x^2 + u_y^2)\frac{\partial^2}{\partial u_+^2}\right)
\end{align}
and we can immediately see that solutions (\ref{CasimirEigenfunction1}) are eigenfunctions of the Casimir operator
\begin{equation}
W^2 \cdot \Phi(u) = (\alpha^2+\beta^2) \Phi(u) = w^2\Phi(u)\,.
\end{equation}
As expected from the properties of the little group, the eigenvalues of the Casimir $W^2$ are non-negative. Analogous solutions were obtained in \cite{Schuster:2013pxj} in examining a scalar master field as a wave function of the continuous spin particle. The analysis here tells us that, due to gauge invariance (\ref{linearHermiteSymmetry}), having the (frame) vector index on a master field does not change the result for the form obtained in \cite{Schuster:2013pxj} for the scalar master field.

A complete orthonormal basis in the auxiliary space can be built from functions of the form (\ref{CasimirEigenfunction1}) for a specific choice of the standard momentum. One possibility is to define
\begin{equation}\label{Basis_Casimir_Eigenfunctions}
f_{\alpha \beta nl}(u) =\frac{1}{\sqrt{2\pi^2}}\exp\left({i\frac{\beta u_x - \alpha u_y}{\omega u_-}}\right)h_n(\omega u_-)h_l (\omega u_-u^2)\,,
\end{equation}
where $h_n(x)$ are any orthonormal and complete functions defined on $\mathbb{R}$, such as Hermite functions.

For $\alpha = \beta = 0$, one has particle states with the vanishing eigenvalue of the second Casimir, which means that they belong to standard massless IRREP in the Wigner classification. We now see explicitly how such states appear inside the continuous spectrum that is parametrised by the eigenvalue of the second Casimir $w^2 = \alpha^2 + \beta^2$.

We prove that the functions $f_{\alpha \beta nl}(u)$ are orthonormal:
\begin{align}
\int d^4u\, f_{\alpha'\beta'n'l'}(u)^*\, f_{\alpha \beta nl}(u) = \frac{1}{(2\pi)^2} &\int_{-\infty}^\infty du_-\, h_{n'}(\omega u_-)^*\, h_n(\omega u_-)\\
\times&\int_{-\infty}^\infty du_1 \int_{-\infty}^\infty du_2\, e^{-i\frac{(\alpha-\alpha') u_2 - (\beta-\beta') u_1}{\omega u_-}}
\nonumber \\ 
\qquad\qquad \times &\int_{-\infty}^\infty du_+\, h_{l'}(\omega\, u_- u^2)^*\, h_l(\omega u_- u^2)
\label{onorm}\,.
\end{align}
We can use a substitution
\begin{equation}
w \equiv \omega\, u_+ u^2 = \omega\, u_+ (-u_+ u_- + u_1^2 + u_2^2)
\end{equation}
to write the third integral as
\begin{equation}
\int_{-\infty}^\infty du_+\, h_{l'}(\omega\, u_+ u^2)^*\, h_l(\omega\, u_+ u^2) = \frac{1}{\omega (u_-)^2} \int_{-\infty}^\infty dw\, h_{l'}(w)^*\, h_l(w)
= \frac{\delta_{l'l}}{\omega (u_-)^2}\,.
\end{equation}
The second integral gives
\begin{equation}
\int_{-\infty}^\infty du_1 
\int_{-\infty}^\infty du_2\, e^{-i\frac{(\alpha-\alpha') u_2 - (\beta-\beta') u_1 }{\omega u_-}} = \left(2\pi \omega u_- \right)^2 \delta(\alpha'-\alpha)\, \delta(\beta'-\beta)
\end{equation}
and in the first integral, we have
\begin{equation}
\int_{-\infty}^\infty du_-\, h_{n'}(\omega u_-)^*\, h_n(\omega u_-) = \frac{1}{\omega^2}\delta_{nn'}\,.
\end{equation}
Finally, we confirm that the basis functions are orthonormal
\begin{equation}
\int d^4u\, f_{\alpha'\beta'n'l'}(u)^*\, f_{\alpha \beta nl}(u) = \delta(\alpha'-\alpha)\, \delta(\beta'-\beta) \,\delta_{l'l}\, \delta_{n'n}\,.
\end{equation}
We can also prove that the choice (\ref{Basis_Casimir_Eigenfunctions}) is complete
\begin{align}
\sum_{n=0}^\infty \sum_{l=0}^\infty \int_{-\infty}^\infty d\alpha  \int_{-\infty}^\infty d \beta\,  f_{\alpha \beta nl}(u')^*\, f_{\alpha \beta nl}(u) =
\frac{1}{2\pi^2} \sum_l h_l(\omega u_-' u^{\prime 2})^*\, h_l(\omega u_- u^2)\nonumber\\
\times \int_{-\infty}^\infty d \alpha  \int_{-\infty}^\infty d\beta\, e^{i\frac{\alpha (u_2'-u_2)-\beta (u_1'-u_1)}{\omega\, u_-}}
\sum_n h_n(\omega u'_-)^*\, h_n(\omega u_-)\label{compr}\,.
\end{align}
Elementary functions such as Hermite satisfy completeness relations
\begin{align}
&\sum_{n=0}^\infty h_n(\omega u'_-)^*\, h_n(\omega u_-) = \delta(\omega(u'_--u_-)) = \frac{1}{|\omega|}\, \delta(u_--u'_-)\\
&\sum_{l=0}^\infty h_l(\omega u_-' u^{\prime 2})^*\, h_l(\omega u_- u^2) = \delta(\omega u_- u^2 - \omega u_-' u^{\prime 2})\,.
\end{align}
With the exponential functions, we have
\begin{align}
&\int_{-\infty}^\infty d\beta\, e^{i\frac{\beta (u_1-u'_1)}{\omega\, u_-}} = 2\pi |\omega u_-| \delta(u_1-u'_1)\\
&\int_{-\infty}^\infty d\alpha\, e^{i\frac{\alpha (u'_2-u_2)}{\omega\, u_-}} = 2\pi |\omega u_-|\delta(u_2-u'_2)\,.
\end{align}
We can insert the results into (\ref{compr}) and obtain
\begin{align}
\sum_n \sum_l &\int d\alpha  \int d\beta\,  f_{\alpha \beta nl}(u')^*\, f_{\alpha \beta nl}(u) =\\=
& 2 \omega (u_-)^2\, \delta(u_--u'_-)\, \delta(u_1-u'_1)\, \delta(u_2-u'_2)\,
\delta(\omega u_- u^2 - \omega u_-' u^{\prime 2})
\nonumber \\
=& 2\, \delta(u_--u'_-)\, \delta(u_1-u'_1)\, \delta(u_2-u'_2)\, \omega (u_-)^2\, \delta(\omega(u_-)^2 (u'_+-u_+))
\nonumber \\
=& 2\, \delta(u_--u'_-)\, \delta(u_1-u'_1)\, \delta(u_2-u'_2)\, \delta(u'_+-u_+)
\nonumber \\
=& \delta^4(u-u')\,,
\end{align}
which is the completeness relation.

The indices $n,l$ are little-group invariant, as well as $\alpha^2+\beta^2$, so we conclude that the content of the massless theory is two polarisations $\times$ infinite $\times$ infinite number of a continuous number of infinite-spin particles. On shell, a classical solution corresponding to a non-vanishing value of the quartic Casimir $W^2 = \alpha^2 + \beta^2$ can be chosen as e.g.
\begin{equation}
h^\pm_{a(\alpha\beta n l k)}(x,u) = \epsilon^\pm_a f_{\alpha \beta nl}(k,u) e^{ikx}
\,,
\end{equation}
where we have emphasised that the polarisation functions have an implicit dependence on the momentum $k^\mu$. To construct a field variable which is square integrable in the auxiliary space, we can form a superposition such as
\begin{align}
h_{a(nlk)}^{\pm}(x,u)=\int\, d \alpha\, d \beta \; c(\alpha,\beta) \, h^\pm_{a(\alpha\beta n l k)}(x,u) \,,
\end{align}
where $\int\, d \alpha\, d \beta \; |c(\alpha,\beta)|^2 < \infty$. Such a superposition goes over various values of $w^2 = \alpha^2+\beta^2$, reminiscent of unparticle physics \cite{Georgi:2007ek} where fields can be thought of as having a continuous distribution of mass \cite{Krasnikov:2007fs, Nikolic:2008ax}.
\section{Discussion}

We explore the particle content of massive and massless free theories of a master field, assuming the requirement that it is square integrable in the auxiliary space. The motivation for such a requirement is that: 1) it ensures a well-defined integration over the auxiliary space without using a nonstandard measure of integration that could violate MHS gauge invariance, 2) it ensures that auxiliary space integration gives a finite value which is required to have finiteness of observables such as energy, 3) it ensures the unitarity of the representation of the part of the Lorentz group that acts on the auxiliary space.
The consequence is that it leads to nonstandard methods of unpacking the spacetime content of the master field. We examine various choices of bases in the auxiliary space where the coefficients of expansion serve as the usual spacetime fields. The technical difficulty is that the representations of the Lorentz group are infinite-dimensional. The result is that in the massless case, the theory contains continuous spin particles in different Poincaré representations specified by two discrete and one continuous label (the continuous being the eigenvalue of the quartic Casimir of the Poincaré group).
In the massive case, the theory contains an infinite number of towers (specified by two discrete labels) of particles containing all spins. 

\begin{acknowledgments}{The work in this paper is based on unpublished research initially presented in the doctoral dissertation of M.P. \cite{Paulisic:2023lam} and extends beyond with
	further investigations and expansions. We thank Loriano Bonora as this work evolved from our mutual collaboration.

The research of P.D.P.\ has been supported by the University of Rijeka under the project uniri-prirod-18-256 and uniri-iskusni-prirod-23-222. The research of M.P.\ has been supported by the University of Rijeka under the project uniri-mladi-prirod-23-43 3159. 
The research of S.G.\ has been supported by a BIRD-2021 project (PRD-2021), the  PRIN Project n.~2022ABPBEY, ``Understanding quantum field theory through its deformations'' and the PRIN 2022 project CONTRABASS (contract n.2022KB2JJM).}
\end{acknowledgments}

\appendix

\section{Massive case}
\label{AppMassive}
\subsection[\appendixname~\thesection]{Particle spectrum}Let us now study the case of massive MHS fields. Apart from a massive MHS matter sector, one can also provide mass to the MHS potential $h_a(x,u)$ by coupling it to the Higgs field in the standard fashion (see \cite{Bonora:2018ggh}). We shall use the simplest case of a scalar field for the purpose of a detailed demonstration, and then generalise the results to vector (and tensor) MHS fields
\begin{equation}
S_0[\varphi] = \frac{1}{2} \int d^4x\, d^4u \left( \partial^x_\mu \varphi(x,u)\,  \partial_x^\mu \varphi(x,u) - m^2\, \varphi(x,u)^2 \right)\,.
\end{equation}

For the moment, there is no difference in the treatment of the massive and massless cases. If we expand the MHS master field by using a complete orthonormal basis in the $u$-space $\{f_r(u)\}$ consisting of real square-integrable functions (say, by using $d$-dimensional Hermite functions) 
\begin{equation} \label{scexp}
\varphi(x,u) = \sum_r \varphi_r(x)\, f_r(u)
\end{equation}
and integrate over $u$, we obtain the following purely spacetime action
\begin{equation}
S_0[\varphi] = \sum_r \frac{1}{2} \int d^4x \left( \partial_\mu \varphi_r(x) \partial^\mu \varphi_r(x)
- m^2 (\varphi_r(x)^2 \right)\,.
\end{equation}
We have obtained a purely spacetime formulation of the theory (which we can extend to the interacting regime) which takes the form of an (infinite) collection of free Klein-Gordon fields all having the same mass $m$. Using this observation we can quantise the theory in the usual way, following the prescription for quantising a set of independent free Klein-Gordon fields to write the quantised spacetime fields  as
\begin{equation} \label{fqKG}
\varphi_r(x) = \int \frac{d^3\mathbf{p}}{\sqrt{(2\pi)^3 \omega_\mathbf{p}}} \left( a_r(p)\,  e^{-i p\cdot x}
+ a_r(p)^\dagger\, e^{i p\cdot x} \right) \quad,\quad p^0 = \omega_\mathbf{p} \equiv \sqrt{\mathbf{p}^2 + m^2}\,.
\end{equation}
Here $a_r(p)^\dagger$ ($a_r(p)$) is the creation (destruction) operator of the type-$r$ quasi-particle carrying 4-momentum $p^\mu$. These operators can be used to construct the Hilbert space of states in the form of the Fock space. E.g., one-particle states are 
\begin{equation} \label{1-part}
| p\, ; r \rangle \equiv a_r(p)^\dagger | 0 \rangle  \quad,\quad
\langle q\, ; r | p\, ; s \rangle = \omega_\mathbf{p} \delta_{rs}\, \delta^3(\mathbf{q} - \mathbf{p})\,,
\end{equation}
where the vacuum state is defined in the usual way
\begin{equation} \label{qpvac}
a_r(p) | 0 \rangle = 0\,.
\end{equation}
Different choices for the orthonormal basis lead to the unitarily equivalent descriptions with the same vacuum state and, correspondingly, the same Hilbert space of states.

Since the fields $\varphi_r(x)$ transform (analogously to (\ref{ha_under_Loretnz})) as
\begin{equation} \label{Ltqsf}
\varphi'_r(x) = U(\Lambda) \varphi_r(x) U(\Lambda)^\dagger =  \sum_s D_{rs}(\Lambda^{-1}) \varphi_s(\Lambda x)\,,
\end{equation}
using (\ref{fqKG}) we see that the quasi-particle states transform under Lorentz transformations as
\begin{eqnarray}
U(\Lambda) | p\, ; r \rangle &=& \sum_s D_{rs}(\Lambda^{-1})\,  | \Lambda p\, ; s \rangle
\nonumber \\
&=& \sum_s D_{sr}(\Lambda)\,  | \Lambda p\, ; s \rangle
\label{Ltrqp}\,.
\end{eqnarray}
As noted in the main text, matrices $D(\Lambda)$ constitute an infinite dimensional unitary representation of the Lorentz group. 
We also note that such a representation is reducible (since if one uses $d$-dimensional Hermite functions
\begin{equation}
f_r(u) = h_{\{n_\mu\}}(u) \equiv h_{n_0}(u_0)\, h_{n_1}(u_1)\, h_{n_2}(u_1)\, h_{n_3}(u_1) \quad,\quad n_\mu = 0,1,2,\ldots
\end{equation}
as a basis, then the subspaces with $N = n_1 + n_2 + n_3 - n_0$ form sub-representations of the Lorentz group).
This is in contrast to the idea that particle-type designation should be a Lorentz-invariant designation. 

As shown above in the section \ref{Wignerology}, it is equivalent to examining either the polarisation functions of solutions to the equations of motion or directly the one-particle states to characterise the particle content of a theory. In the appendix, we choose to work with one-particle states. Therefore we also note that different types of bases used in the main text correspond to different types of particles in the following sense: a momentum-independent basis, such as the product of Hermite polynomials, gives rise to one-particle states that we refer to as quasi-particle states since they do not possess a Lorentz-invariant designation, and that the choice of a momentum dependent basis such as $g_{n_0 n s \sigma}(u)$ (see below) gives rise to one-particle states which do posses a Lorentz-invariant designation.

To analyse the content of the one-particle sector, spanned by (\ref{1-part}), as in the massless case, we need to write it as a direct sum of IRREPs of the Poincaré group by using Wigner's little group construction. 
In effect, we need to obtain polarisation functions (i.e.\ the basis functions in the auxiliary space) appropriate for the massive case. 
First, we choose a special 4-momentum tuned to the massive case
\begin{equation} \label{mchm}
k^\mu = (m,0,0,0) \,.
\end{equation}
The subgroup of the Lorentz group, which keeps $k$ invariant (little group), is the $SO(3)$ group of rotations. We know that IRREPs of the Poincaré group, in this case, are classified by the spin $s=0,1,2,\ldots$, which denote (unique) $(2s+1)$-dimensional representations. 

Since the basis states transform as in (\ref{Ltrqp}), the problem can be transcribed into the problem of diagonalising the action of rotations, given by 
\begin{equation} \label{Ltbf}
f_r(\Lambda^{-1} u) = \sum_s D_{rs}(\Lambda^{-1}) f_s(u)
\end{equation}
over the space of $L_2(\mathbb{R}^4)$ functions on the auxiliary space. But this is a well-known problem whose solution is to take the basis built upon the spherical harmonics. For example, one can choose the orthonormal basis in the following way
\begin{equation}
g_{n_0 n s \sigma}(u) = h_{n_0}(u_0) R_{ns}(u_r) e^{-u_r^2/2} u_r^s Y^\sigma_s(u_\theta,u_\phi)\,,
\end{equation}
where $u_0$, $u_r$, $u_\theta$ and $u_\phi$ are the spherical coordinates of the auxiliary space, $\{h_{n_0}, n_0=1,2,\ldots\}$ are Hermite functions, $\{Y_l^m, l=0,1,\ldots; m=-l,\ldots,l\}$ are spherical harmonics, and $\{R_{nl}, n=0,1,2,\ldots\}$ are real polynomials of the order $2n$ satisfying
\begin{equation}
\int_0^\infty du_r u_r^{2(l+1)} e^{-u_r^2} R_{nl}(u_r) R_{n'l}(u_r) = \delta_{nn'} 
\end{equation}
and the corresponding completeness condition. Note that this basis is not real
\begin{equation}
g_{n_0 n s \sigma}(u)^* = g_{n_0 n s -\sigma}(u)\,.
\end{equation}
This basis transforms under $SO(3)$ rotations as
\begin{equation}
g_{n_0 n s \sigma}(R^{-1} u) = \mathcal{D}_{\sigma \sigma'}^{(s)}(R^{-1})\, g_{n_0 n s \sigma'}(u)
\end{equation}
where $\mathcal{D}_{\sigma \sigma'}^{(s)}(R)$ are the usual spin-$s$ rotation matrices. If we define creation operators 
$a_{n_0 n s \sigma}(k)^\dagger$ by
\begin{equation}
\sum_r f_r(u)\, a_r(k)^\dagger = \sum_{n_0=0}^\infty \sum_{n=0}^\infty \sum_{s=0}^\infty \sum_{\sigma=-s}^s 
g_{n_0 n s \sigma}(u)\, a_{n_0 n s \sigma}(k)^\dagger
\end{equation}
then they transform under the rotations as
\begin{equation}
U(R) a_{n_0 n s \sigma}(k)^\dagger U(R)^\dagger = \sum_{\sigma'=-s}^s
\mathcal{D}_{\sigma' \sigma}^{(s)}(R)\, a_{n_0 n s \sigma'}(k)^\dagger \,.
\end{equation}
The vacuum and one-particle states carrying momentum $p=k$ given in (\ref{mchm}) are defined by
\begin{equation}
| k , \sigma; n_0, n, s \rangle = a_{n_0 n s \sigma}(k)^\dagger | 0 \rangle \qquad,\qquad
a_{n_0 n s \sigma}(k) | 0 \rangle = 0\,.
\end{equation}
These states transform under rotations as
\begin{equation}
U(R) | k , \sigma; n_0, n, s \rangle = \sum_{\sigma'=-s}^s \mathcal{D}_{\sigma' \sigma}^{(s)}(R)\, | k , \sigma'; n_0, n, s \rangle\,,
\end{equation}
from which it follows that
\begin{equation}
\mathbf{J}^2 | k , \sigma; n_0, n, s \rangle = s(s+1) | k , \sigma; n_0, n, s \rangle \quad,\quad
J_z | k , \sigma; n_0, n, s \rangle = \sigma | k , \sigma; n_0, n, s \rangle\,,
\end{equation}
which means that they describe particle states of spin $s$. The one-particle subspace with momentum $k$ is the direct sum of particles labelled by the triplet of numbers $(n_0,n,s)$ with $n_0, n, s = 0,1,2,\ldots$\,.

To complete the description we must construct one-particle states with generic on-shell momentum $p$, which can be done in the standard fashion
\begin{equation} \label{Ltrp}
| p , \sigma; n_0, n, s \rangle \equiv U(\Lambda(p)) | k , \sigma; n_0, n, s \rangle \quad,\quad p = \Lambda(p) k\,,
\end{equation}
where $\Lambda(p)$ is a pure boost. 

In summary, the massive theory contains an infinite number of standard massive particles labelled by the (little group invariant) triplet of numbers $(n_0,n,s)$ with $n_0, n, s = 0,1,2,\ldots$. We have an infinite $\times$ infinite number of towers of particles containing all spins.

\subsubsection[\appendixname~\thesection]{Higher-tensor massive case}

What if we have an MHS master field which is not scalar? Let us use a massive vector MHS master field $h_a(x,u)$ as an example (it could be the MHS potential if the MHS symmetry is spontaneously broken.). Using the expansion
\begin{equation}
h^a(x,u) = \sum_r h^a_r(x)\, f_r(u)\,,
\end{equation}
the free field theory boils down to a collection of spacetime fields $h^a_r(x)$ all satisfying the Proca equation with the same mass $m$. These fields transform under the Lorentz group as
\begin{equation}
h^{a\prime}_r(x) =  (\Lambda)^a{}_b \sum_s D_{rs}(\Lambda) h^b_s(\Lambda^{-1} x)\,.
\end{equation}
By quantising Proca spacetime fields in the usual way one gets the one particle spectrum consisting of states
\begin{equation}
| p , \sigma_1 ; r \rangle\,,
\end{equation}
where $\sigma_1 = -1, 0, 1$ is the index belonging to the spin-1 unitary IRREP of $SO(3)$ little group. The whole procedure applied for the scalar master field can be repeated here with the only difference that when block-diagonalising rotation matrices, we have to take into account that spin coming from auxiliary space is here multiplied with the spin-1 state coming from the vectorial property of the master field. Here one uses the standard formula for a direct product of two $SO(3)$ unitary IRREPs to obtain that in the classification of Lorentz particles, instead of a single particle of spin $s$ we will have particles of spin
\begin{equation} \label{tmhsfe}
1 \otimes s = (s+1) \oplus s \oplus (s-1) \;,\quad s\ge1 \qquad\mbox{and}\qquad 1 \otimes 0 = 1\,.
\end{equation}
The conclusion is that for every particle of spin $s=j$ present in the spectrum of the free massive \textit{scalar} master field, one obtains a ``triplet'' of particles with the spins $s=j+1,j,j-1$ for $j\ge1$, and $s=0,1$ for $j=0$. The result has an obvious generalisation to higher-rank tensor master fields.

\bibliographystyle{apsrev4-1}
\bibliography{references.bib}

\begin{thebibliography}{29}%
\makeatletter
\providecommand \@ifxundefined [1]{%
 \@ifx{#1\undefined}
}%
\providecommand \@ifnum [1]{%
 \ifnum #1\expandafter \@firstoftwo
 \else \expandafter \@secondoftwo
 \fi
}%
\providecommand \@ifx [1]{%
 \ifx #1\expandafter \@firstoftwo
 \else \expandafter \@secondoftwo
 \fi
}%
\providecommand \natexlab [1]{#1}%
\providecommand \enquote  [1]{``#1''}%
\providecommand \bibnamefont  [1]{#1}%
\providecommand \bibfnamefont [1]{#1}%
\providecommand \citenamefont [1]{#1}%
\providecommand \href@noop [0]{\@secondoftwo}%
\providecommand \href [0]{\begingroup \@sanitize@url \@href}%
\providecommand \@href[1]{\@@startlink{#1}\@@href}%
\providecommand \@@href[1]{\endgroup#1\@@endlink}%
\providecommand \@sanitize@url [0]{\catcode `\\12\catcode `\$12\catcode
  `\&12\catcode `\#12\catcode `\^12\catcode `\_12\catcode `\%12\relax}%
\providecommand \@@startlink[1]{}%
\providecommand \@@endlink[0]{}%
\providecommand \url  [0]{\begingroup\@sanitize@url \@url }%
\providecommand \@url [1]{\endgroup\@href {#1}{\urlprefix }}%
\providecommand \urlprefix  [0]{URL }%
\providecommand \Eprint [0]{\href }%
\providecommand \doibase [0]{http://dx.doi.org/}%
\providecommand \selectlanguage [0]{\@gobble}%
\providecommand \bibinfo  [0]{\@secondoftwo}%
\providecommand \bibfield  [0]{\@secondoftwo}%
\providecommand \translation [1]{[#1]}%
\providecommand \BibitemOpen [0]{}%
\providecommand \bibitemStop [0]{}%
\providecommand \bibitemNoStop [0]{.\EOS\space}%
\providecommand \EOS [0]{\spacefactor3000\relax}%
\providecommand \BibitemShut  [1]{\csname bibitem#1\endcsname}%
\let\auto@bib@innerbib\@empty
\bibitem [{\citenamefont {Cvitan}\ \emph
  {et~al.}(2021{\natexlab{a}})\citenamefont {Cvitan}, \citenamefont
  {Dominis~Prester}, \citenamefont {Giaccari}, \citenamefont {Pauli\v{s}i\'c},\
  and\ \citenamefont {Vukovi\'c}}]{Cvitan:2021yvf}%
  \BibitemOpen
  \bibfield  {author} {\bibinfo {author} {\bibfnamefont {M.}~\bibnamefont
  {Cvitan}}, \bibinfo {author} {\bibfnamefont {P.}~\bibnamefont
  {Dominis~Prester}}, \bibinfo {author} {\bibfnamefont {S.}~\bibnamefont
  {Giaccari}}, \bibinfo {author} {\bibfnamefont {M.}~\bibnamefont
  {Pauli\v{s}i\'c}}, \ and\ \bibinfo {author} {\bibfnamefont {I.}~\bibnamefont
  {Vukovi\'c}},\ }\href {\doibase 10.1007/JHEP06(2021)144} {\bibfield
  {journal} {\bibinfo  {journal} {JHEP}\ }\textbf {\bibinfo {volume} {06}},\
  \bibinfo {pages} {144} (\bibinfo {year} {2021}{\natexlab{a}})},\ \Eprint
  {http://arxiv.org/abs/2102.09254} {arXiv:2102.09254 [hep-th]} \BibitemShut
  {NoStop}%
\bibitem [{\citenamefont {Cvitan}\ \emph
  {et~al.}(2021{\natexlab{b}})\citenamefont {Cvitan}, \citenamefont {Prester},
  \citenamefont {Giaccari}, \citenamefont {Pauli\v{s}i\'c},\ and\ \citenamefont
  {Vukovi\'c}}]{Cvitan:2021qvm}%
  \BibitemOpen
  \bibfield  {author} {\bibinfo {author} {\bibfnamefont {M.}~\bibnamefont
  {Cvitan}}, \bibinfo {author} {\bibfnamefont {P.~D.}\ \bibnamefont {Prester}},
  \bibinfo {author} {\bibfnamefont {S.~G.}\ \bibnamefont {Giaccari}}, \bibinfo
  {author} {\bibfnamefont {M.}~\bibnamefont {Pauli\v{s}i\'c}}, \ and\ \bibinfo
  {author} {\bibfnamefont {I.}~\bibnamefont {Vukovi\'c}},\ }\href {\doibase
  10.3390/sym13091581} {\bibfield  {journal} {\bibinfo  {journal} {Symmetry}\
  }\textbf {\bibinfo {volume} {13}},\ \bibinfo {pages} {1581} (\bibinfo {year}
  {2021}{\natexlab{b}})}\BibitemShut {NoStop}%
\bibitem [{\citenamefont {Cvitan}\ \emph {et~al.}(2022)\citenamefont {Cvitan},
  \citenamefont {Dominis~Prester}, \citenamefont {Giaccari}, \citenamefont
  {Pauli\v{s}i\'c},\ and\ \citenamefont {Vukovi\'c}}]{Cvitan:2022wzf}%
  \BibitemOpen
  \bibfield  {author} {\bibinfo {author} {\bibfnamefont {M.}~\bibnamefont
  {Cvitan}}, \bibinfo {author} {\bibfnamefont {P.}~\bibnamefont
  {Dominis~Prester}}, \bibinfo {author} {\bibfnamefont {S.~G.}\ \bibnamefont
  {Giaccari}}, \bibinfo {author} {\bibfnamefont {M.}~\bibnamefont
  {Pauli\v{s}i\'c}}, \ and\ \bibinfo {author} {\bibfnamefont {I.}~\bibnamefont
  {Vukovi\'c}},\ }\href {\doibase 10.1007/978-981-19-4751-3_43} {\bibfield
  {journal} {\bibinfo  {journal} {Springer Proc. Math. Stat.}\ }\textbf
  {\bibinfo {volume} {396}},\ \bibinfo {pages} {463} (\bibinfo {year}
  {2022})}\BibitemShut {NoStop}%
\bibitem [{\citenamefont {Schuster}\ and\ \citenamefont
  {Toro}(2013{\natexlab{a}})}]{Schuster:2013pta}%
  \BibitemOpen
  \bibfield  {author} {\bibinfo {author} {\bibfnamefont {P.}~\bibnamefont
  {Schuster}}\ and\ \bibinfo {author} {\bibfnamefont {N.}~\bibnamefont
  {Toro}},\ }\href {\doibase 10.1007/JHEP10(2013)061} {\bibfield  {journal}
  {\bibinfo  {journal} {JHEP}\ }\textbf {\bibinfo {volume} {10}},\ \bibinfo
  {pages} {061} (\bibinfo {year} {2013}{\natexlab{a}})},\ \Eprint
  {http://arxiv.org/abs/1302.3225} {arXiv:1302.3225 [hep-th]} \BibitemShut
  {NoStop}%
\bibitem [{\citenamefont {Bonora}\ \emph
  {et~al.}(2018{\natexlab{a}})\citenamefont {Bonora}, \citenamefont {Cvitan},
  \citenamefont {Dominis~Prester}, \citenamefont {Giaccari}, \citenamefont
  {Pauli\v{s}i\'c},\ and\ \citenamefont {\v{S}temberga}}]{Bonora:2018uwx}%
  \BibitemOpen
  \bibfield  {author} {\bibinfo {author} {\bibfnamefont {L.}~\bibnamefont
  {Bonora}}, \bibinfo {author} {\bibfnamefont {M.}~\bibnamefont {Cvitan}},
  \bibinfo {author} {\bibfnamefont {P.}~\bibnamefont {Dominis~Prester}},
  \bibinfo {author} {\bibfnamefont {S.}~\bibnamefont {Giaccari}}, \bibinfo
  {author} {\bibfnamefont {M.}~\bibnamefont {Pauli\v{s}i\'c}}, \ and\ \bibinfo
  {author} {\bibfnamefont {T.}~\bibnamefont {\v{S}temberga}},\ }\href {\doibase
  10.1007/JHEP04(2018)095} {\bibfield  {journal} {\bibinfo  {journal} {JHEP}\
  }\textbf {\bibinfo {volume} {04}},\ \bibinfo {pages} {095} (\bibinfo {year}
  {2018}{\natexlab{a}})},\ \Eprint {http://arxiv.org/abs/1802.02968}
  {arXiv:1802.02968 [hep-th]} \BibitemShut {NoStop}%
\bibitem [{\citenamefont {Bonora}\ \emph {et~al.}(2019)\citenamefont {Bonora},
  \citenamefont {Cvitan}, \citenamefont {Dominis~Prester}, \citenamefont
  {Giaccari},\ and\ \citenamefont {\v{S}temberga}}]{Bonora:2018eot}%
  \BibitemOpen
  \bibfield  {author} {\bibinfo {author} {\bibfnamefont {L.}~\bibnamefont
  {Bonora}}, \bibinfo {author} {\bibfnamefont {M.}~\bibnamefont {Cvitan}},
  \bibinfo {author} {\bibfnamefont {P.}~\bibnamefont {Dominis~Prester}},
  \bibinfo {author} {\bibfnamefont {S.}~\bibnamefont {Giaccari}}, \ and\
  \bibinfo {author} {\bibfnamefont {T.}~\bibnamefont {\v{S}temberga}},\ }\href
  {\doibase 10.1140/epjc/s10052-019-6660-4} {\bibfield  {journal} {\bibinfo
  {journal} {Eur. Phys. J. C}\ }\textbf {\bibinfo {volume} {79}},\ \bibinfo
  {pages} {258} (\bibinfo {year} {2019})},\ \Eprint
  {http://arxiv.org/abs/1811.04847} {arXiv:1811.04847 [hep-th]} \BibitemShut
  {NoStop}%
\bibitem [{\citenamefont {Bonora}\ \emph
  {et~al.}(2018{\natexlab{b}})\citenamefont {Bonora}, \citenamefont {Cvitan},
  \citenamefont {Dominis~Prester}, \citenamefont {Giaccari},\ and\
  \citenamefont {Stemberga}}]{Bonora:2018ggh}%
  \BibitemOpen
  \bibfield  {author} {\bibinfo {author} {\bibfnamefont {L.}~\bibnamefont
  {Bonora}}, \bibinfo {author} {\bibfnamefont {M.}~\bibnamefont {Cvitan}},
  \bibinfo {author} {\bibfnamefont {P.}~\bibnamefont {Dominis~Prester}},
  \bibinfo {author} {\bibfnamefont {S.}~\bibnamefont {Giaccari}}, \ and\
  \bibinfo {author} {\bibfnamefont {T.}~\bibnamefont {Stemberga}},\ }\href@noop
  {} {\  (\bibinfo {year} {2018}{\natexlab{b}})},\ \Eprint
  {http://arxiv.org/abs/1812.05030} {arXiv:1812.05030 [hep-th]} \BibitemShut
  {NoStop}%
\bibitem [{\citenamefont {Bonora}(2018)}]{Bonora:2018mqg}%
  \BibitemOpen
  \bibfield  {author} {\bibinfo {author} {\bibfnamefont {L.}~\bibnamefont
  {Bonora}},\ }\href {\doibase 10.1142/S0217751X18450070} {\bibfield  {journal}
  {\bibinfo  {journal} {Int. J. Mod. Phys. A}\ }\textbf {\bibinfo {volume}
  {33}},\ \bibinfo {pages} {1845007} (\bibinfo {year} {2018})}\BibitemShut
  {NoStop}%
\bibitem [{\citenamefont {Bonora}\ and\ \citenamefont
  {Giaccari}(2020)}]{Bonora:2020aqp}%
  \BibitemOpen
  \bibfield  {author} {\bibinfo {author} {\bibfnamefont {L.}~\bibnamefont
  {Bonora}}\ and\ \bibinfo {author} {\bibfnamefont {S.}~\bibnamefont
  {Giaccari}},\ }\href {\doibase 10.3390/universe6120245} {\bibfield  {journal}
  {\bibinfo  {journal} {Universe}\ }\textbf {\bibinfo {volume} {6}},\ \bibinfo
  {pages} {245} (\bibinfo {year} {2020})},\ \Eprint
  {http://arxiv.org/abs/2011.00734} {arXiv:2011.00734 [hep-th]} \BibitemShut
  {NoStop}%
\bibitem [{\citenamefont {Schuster}\ and\ \citenamefont
  {Toro}(2013{\natexlab{b}})}]{Schuster:2013pxj}%
  \BibitemOpen
  \bibfield  {author} {\bibinfo {author} {\bibfnamefont {P.}~\bibnamefont
  {Schuster}}\ and\ \bibinfo {author} {\bibfnamefont {N.}~\bibnamefont
  {Toro}},\ }\href {\doibase 10.1007/JHEP09(2013)104} {\bibfield  {journal}
  {\bibinfo  {journal} {JHEP}\ }\textbf {\bibinfo {volume} {09}},\ \bibinfo
  {pages} {104} (\bibinfo {year} {2013}{\natexlab{b}})},\ \Eprint
  {http://arxiv.org/abs/1302.1198} {arXiv:1302.1198 [hep-th]} \BibitemShut
  {NoStop}%
\bibitem [{\citenamefont {Cvitan}\ \emph {et~al.}(2024)\citenamefont {Cvitan},
  \citenamefont {Dominis~Prester}, \citenamefont {Giaccari}, \citenamefont
  {Pauli\v{s}i\'c},\ and\ \citenamefont {Vukovi\'c}}]{Cvitan:2024xmn}%
  \BibitemOpen
  \bibfield  {author} {\bibinfo {author} {\bibfnamefont {M.}~\bibnamefont
  {Cvitan}}, \bibinfo {author} {\bibfnamefont {P.}~\bibnamefont
  {Dominis~Prester}}, \bibinfo {author} {\bibfnamefont {S.}~\bibnamefont
  {Giaccari}}, \bibinfo {author} {\bibfnamefont {M.}~\bibnamefont
  {Pauli\v{s}i\'c}}, \ and\ \bibinfo {author} {\bibfnamefont {I.}~\bibnamefont
  {Vukovi\'c}},\ }\href@noop {} {\  (\bibinfo {year} {2024})},\ \Eprint
  {http://arxiv.org/abs/2405.00404} {arXiv:2405.00404 [hep-th]} \BibitemShut
  {NoStop}%
\bibitem [{\citenamefont {Bengtsson}(2008)}]{Bengtsson:2008mw}%
  \BibitemOpen
  \bibfield  {author} {\bibinfo {author} {\bibfnamefont {A.~K.~H.}\
  \bibnamefont {Bengtsson}},\ }\href {\doibase 10.3842/SIGMA.2008.013}
  {\bibfield  {journal} {\bibinfo  {journal} {SIGMA}\ }\textbf {\bibinfo
  {volume} {4}},\ \bibinfo {pages} {013} (\bibinfo {year} {2008})},\ \Eprint
  {http://arxiv.org/abs/0802.0479} {arXiv:0802.0479 [hep-th]} \BibitemShut
  {NoStop}%
\bibitem [{\citenamefont {Weinberg}(2005)}]{Weinberg:1995mt}%
  \BibitemOpen
  \bibfield  {author} {\bibinfo {author} {\bibfnamefont {S.}~\bibnamefont
  {Weinberg}},\ }\href@noop {} {\emph {\bibinfo {title} {{The Quantum theory of
  fields. Vol. 1: Foundations}}}}\ (\bibinfo  {publisher} {Cambridge University
  Press},\ \bibinfo {year} {2005})\BibitemShut {NoStop}%
\bibitem [{\citenamefont {Loebbert}(2008)}]{Loebbert:2008zz}%
  \BibitemOpen
  \bibfield  {author} {\bibinfo {author} {\bibfnamefont {F.}~\bibnamefont
  {Loebbert}},\ }\href {\doibase 10.1002/andp.200810305} {\bibfield  {journal}
  {\bibinfo  {journal} {Annalen Phys.}\ }\textbf {\bibinfo {volume} {17}},\
  \bibinfo {pages} {803} (\bibinfo {year} {2008})}\BibitemShut {NoStop}%
\bibitem [{\citenamefont {Duncan}(2012)}]{Duncan:2012aja}%
  \BibitemOpen
  \bibfield  {author} {\bibinfo {author} {\bibfnamefont {A.}~\bibnamefont
  {Duncan}},\ }\href {\doibase 10.1093/acprof:oso/9780199573264.001.0001}
  {\emph {\bibinfo {title} {{The Conceptual Framework of Quantum Field
  Theory}}}}\ (\bibinfo  {publisher} {Oxford University Press},\ \bibinfo
  {year} {2012})\BibitemShut {NoStop}%
\bibitem [{\citenamefont {Bekaert}\ and\ \citenamefont
  {Boulanger}(2021)}]{Bekaert:2006py}%
  \BibitemOpen
  \bibfield  {author} {\bibinfo {author} {\bibfnamefont {X.}~\bibnamefont
  {Bekaert}}\ and\ \bibinfo {author} {\bibfnamefont {N.}~\bibnamefont
  {Boulanger}},\ }\href {\doibase 10.21468/SciPostPhysLectNotes.30} {\bibfield
  {journal} {\bibinfo  {journal} {SciPost Phys. Lect. Notes}\ }\textbf
  {\bibinfo {volume} {30}},\ \bibinfo {pages} {1} (\bibinfo {year} {2021})},\
  \Eprint {http://arxiv.org/abs/hep-th/0611263} {arXiv:hep-th/0611263}
  \BibitemShut {NoStop}%
\bibitem [{\citenamefont {Wigner}(1939)}]{Wigner:1939cj}%
  \BibitemOpen
  \bibfield  {author} {\bibinfo {author} {\bibfnamefont {E.~P.}\ \bibnamefont
  {Wigner}},\ }\href {\doibase 10.2307/1968551} {\bibfield  {journal} {\bibinfo
   {journal} {Annals Math.}\ }\textbf {\bibinfo {volume} {40}},\ \bibinfo
  {pages} {149} (\bibinfo {year} {1939})}\BibitemShut {NoStop}%
\bibitem [{\citenamefont {Tung}(1985)}]{Tung:1985na}%
  \BibitemOpen
  \bibfield  {author} {\bibinfo {author} {\bibfnamefont {W.~K.}\ \bibnamefont
  {Tung}},\ }\href@noop {} {\emph {\bibinfo {title} {{GROUP THEORY IN
  PHYSICS}}}}\ (\bibinfo {year} {1985})\BibitemShut {NoStop}%
\bibitem [{\citenamefont {Schuster}\ and\ \citenamefont
  {Toro}(2013{\natexlab{c}})}]{Schuster:2013vpr}%
  \BibitemOpen
  \bibfield  {author} {\bibinfo {author} {\bibfnamefont {P.}~\bibnamefont
  {Schuster}}\ and\ \bibinfo {author} {\bibfnamefont {N.}~\bibnamefont
  {Toro}},\ }\href {\doibase 10.1007/JHEP09(2013)105} {\bibfield  {journal}
  {\bibinfo  {journal} {JHEP}\ }\textbf {\bibinfo {volume} {09}},\ \bibinfo
  {pages} {105} (\bibinfo {year} {2013}{\natexlab{c}})},\ \Eprint
  {http://arxiv.org/abs/1302.1577} {arXiv:1302.1577 [hep-th]} \BibitemShut
  {NoStop}%
\bibitem [{\citenamefont {Schuster}\ and\ \citenamefont
  {Toro}(2015)}]{Schuster:2014hca}%
  \BibitemOpen
  \bibfield  {author} {\bibinfo {author} {\bibfnamefont {P.}~\bibnamefont
  {Schuster}}\ and\ \bibinfo {author} {\bibfnamefont {N.}~\bibnamefont
  {Toro}},\ }\href {\doibase 10.1103/PhysRevD.91.025023} {\bibfield  {journal}
  {\bibinfo  {journal} {Phys. Rev. D}\ }\textbf {\bibinfo {volume} {91}},\
  \bibinfo {pages} {025023} (\bibinfo {year} {2015})},\ \Eprint
  {http://arxiv.org/abs/1404.0675} {arXiv:1404.0675 [hep-th]} \BibitemShut
  {NoStop}%
\bibitem [{\citenamefont {Bekaert}\ and\ \citenamefont
  {Skvortsov}(2017)}]{Bekaert:2017khg}%
  \BibitemOpen
  \bibfield  {author} {\bibinfo {author} {\bibfnamefont {X.}~\bibnamefont
  {Bekaert}}\ and\ \bibinfo {author} {\bibfnamefont {E.~D.}\ \bibnamefont
  {Skvortsov}},\ }\href {\doibase 10.1142/S0217751X17300198} {\bibfield
  {journal} {\bibinfo  {journal} {Int. J. Mod. Phys. A}\ }\textbf {\bibinfo
  {volume} {32}},\ \bibinfo {pages} {1730019} (\bibinfo {year} {2017})},\
  \Eprint {http://arxiv.org/abs/1708.01030} {arXiv:1708.01030 [hep-th]}
  \BibitemShut {NoStop}%
\bibitem [{\citenamefont {Rivelles}(2017)}]{Rivelles:2016rwo}%
  \BibitemOpen
  \bibfield  {author} {\bibinfo {author} {\bibfnamefont {V.~O.}\ \bibnamefont
  {Rivelles}},\ }\href {\doibase 10.1140/epjc/s10052-017-4927-1} {\bibfield
  {journal} {\bibinfo  {journal} {Eur. Phys. J. C}\ }\textbf {\bibinfo {volume}
  {77}},\ \bibinfo {pages} {433} (\bibinfo {year} {2017})},\ \Eprint
  {http://arxiv.org/abs/1607.01316} {arXiv:1607.01316 [hep-th]} \BibitemShut
  {NoStop}%
\bibitem [{\citenamefont {Schuster}\ \emph {et~al.}(2023)\citenamefont
  {Schuster}, \citenamefont {Toro},\ and\ \citenamefont
  {Zhou}}]{Schuster:2023xqa}%
  \BibitemOpen
  \bibfield  {author} {\bibinfo {author} {\bibfnamefont {P.}~\bibnamefont
  {Schuster}}, \bibinfo {author} {\bibfnamefont {N.}~\bibnamefont {Toro}}, \
  and\ \bibinfo {author} {\bibfnamefont {K.}~\bibnamefont {Zhou}},\ }\href
  {\doibase 10.1007/JHEP04(2023)010} {\bibfield  {journal} {\bibinfo  {journal}
  {JHEP}\ }\textbf {\bibinfo {volume} {04}},\ \bibinfo {pages} {010} (\bibinfo
  {year} {2023})},\ \Eprint {http://arxiv.org/abs/2303.04816} {arXiv:2303.04816
  [hep-th]} \BibitemShut {NoStop}%
\bibitem [{\citenamefont {Schuster}\ and\ \citenamefont
  {Toro}(2024)}]{Schuster:2023jgc}%
  \BibitemOpen
  \bibfield  {author} {\bibinfo {author} {\bibfnamefont {P.}~\bibnamefont
  {Schuster}}\ and\ \bibinfo {author} {\bibfnamefont {N.}~\bibnamefont
  {Toro}},\ }\href {\doibase 10.1103/PhysRevD.109.096008} {\bibfield  {journal}
  {\bibinfo  {journal} {Phys. Rev. D}\ }\textbf {\bibinfo {volume} {109}},\
  \bibinfo {pages} {096008} (\bibinfo {year} {2024})},\ \Eprint
  {http://arxiv.org/abs/2308.16218} {arXiv:2308.16218 [hep-th]} \BibitemShut
  {NoStop}%
\bibitem [{\citenamefont {Schuster}\ \emph {et~al.}(2024)\citenamefont
  {Schuster}, \citenamefont {Sundaresan},\ and\ \citenamefont
  {Toro}}]{Schuster:2024wjc}%
  \BibitemOpen
  \bibfield  {author} {\bibinfo {author} {\bibfnamefont {P.}~\bibnamefont
  {Schuster}}, \bibinfo {author} {\bibfnamefont {G.}~\bibnamefont
  {Sundaresan}}, \ and\ \bibinfo {author} {\bibfnamefont {N.}~\bibnamefont
  {Toro}},\ }\href@noop {} {\  (\bibinfo {year} {2024})},\ \Eprint
  {http://arxiv.org/abs/2406.14616} {arXiv:2406.14616 [hep-ph]} \BibitemShut
  {NoStop}%
\bibitem [{\citenamefont {Georgi}(2007)}]{Georgi:2007ek}%
  \BibitemOpen
  \bibfield  {author} {\bibinfo {author} {\bibfnamefont {H.}~\bibnamefont
  {Georgi}},\ }\href {\doibase 10.1103/PhysRevLett.98.221601} {\bibfield
  {journal} {\bibinfo  {journal} {Phys. Rev. Lett.}\ }\textbf {\bibinfo
  {volume} {98}},\ \bibinfo {pages} {221601} (\bibinfo {year} {2007})},\
  \Eprint {http://arxiv.org/abs/hep-ph/0703260} {arXiv:hep-ph/0703260}
  \BibitemShut {NoStop}%
\bibitem [{\citenamefont {Krasnikov}(2007)}]{Krasnikov:2007fs}%
  \BibitemOpen
  \bibfield  {author} {\bibinfo {author} {\bibfnamefont {N.~V.}\ \bibnamefont
  {Krasnikov}},\ }\href {\doibase 10.1142/S0217751X07037342} {\bibfield
  {journal} {\bibinfo  {journal} {Int. J. Mod. Phys. A}\ }\textbf {\bibinfo
  {volume} {22}},\ \bibinfo {pages} {5117} (\bibinfo {year} {2007})},\ \Eprint
  {http://arxiv.org/abs/0707.1419} {arXiv:0707.1419 [hep-ph]} \BibitemShut
  {NoStop}%
\bibitem [{\citenamefont {Nikolic}(2008)}]{Nikolic:2008ax}%
  \BibitemOpen
  \bibfield  {author} {\bibinfo {author} {\bibfnamefont {H.}~\bibnamefont
  {Nikolic}},\ }\href {\doibase 10.1142/S021773230802820X} {\bibfield
  {journal} {\bibinfo  {journal} {Mod. Phys. Lett. A}\ }\textbf {\bibinfo
  {volume} {23}},\ \bibinfo {pages} {2645} (\bibinfo {year} {2008})},\ \Eprint
  {http://arxiv.org/abs/0801.4471} {arXiv:0801.4471 [hep-ph]} \BibitemShut
  {NoStop}%
\bibitem [{\citenamefont {Pauli\v{s}i\'c}(2023)}]{Paulisic:2023lam}%
  \BibitemOpen
  \bibfield  {author} {\bibinfo {author} {\bibfnamefont {M.}~\bibnamefont
  {Pauli\v{s}i\'c}},\ }\emph {\bibinfo {title} {{Higher-spin-like symmetries
  and gauge models}}},\ \href@noop {} {Ph.D. thesis},\ \bibinfo  {school}
  {University of Rijeka} (\bibinfo {year} {2023}),\ \bibinfo {note}
  {\url{https://urn.nsk.hr/urn:nbn:hr:194:926638}}\BibitemShut {NoStop}%
\end{thebibliography}%
\end{document}